\documentclass[useAMS,usenatbib]{mn2e}
\pdfoutput=1
\usepackage{graphicx}
\usepackage{color}

\title[The SkyMapper DR1.1 Search for Extremely Metal-Poor Stars] 
{The SkyMapper DR1.1 Search for Extremely Metal-Poor Stars}
\author[Da Costa et al.]
{G. S. Da Costa,$^1$ M. S. Bessell,$^1$ A. D. Mackey,$^{1,2}$  T. Nordlander,$^{1,2}$ M. Asplund,$^{1,2}$  
\newauthor
A. R. Casey,$^{1,3,4}$ A. Frebel,$^5$ K. Lind,$^{6,7}$ A. F. Marino,$^{1,8,9}$ S. J. Murphy,$^{1,10}$     
\newauthor
J. E. Norris,$^1$ B. P. Schmidt,$^1$ and D. Yong$^{1,2}$\\ 
\\
$^{1}$Research School of Astronomy and Astrophysics, Australian National 
University, Canberra, ACT 0200, Australia\\
$^{2}$ARC Centre of Excellence for Astrophysics in Three Dimensions (ASTRO-3D), Australia\\
$^{3}$School of Physics and Astronomy, Monash University, Wellington Rd, Clayton, VIC 3800, Australia\\
$^{4}$Faculty of Information Technology, Monash University, Wellington Rd, Clayton, VIC 3800, Australia\\
$^{5}$Department of Physics and Kavli Institute for Astrophysics and Space Research, Massachusetts Institute of Technology,\\
Cambridge, MA 02139, USA\\
$^{6}$Max-Planck-Institut f\"{u}r Astronomie, K\"{o}nigstuhl 17, D-69117 Heidelberg, Germany\\
$^{7}$Observational Astrophysics, Department of Physics and Astronomy, Uppsala University, Box 516, SE-751 20 Uppsala, Sweden\\
$^{8}$Dipartimento di Fisica e Astronomia `Galileo Galilei' -- Univ.\ di Padova, Vicolo dellÕOsservatorio 3, Padova, IT-35122, Italy\\
$^{9}$Centro di Ateneo di Studi e Attivita Spaziali `Giuseppe Colombo' -- CISAS, Via Venezia 15, Padova, IT-35131, Italy\\
$^{10}$School of Science, University of New South Wales, Canberra, ACT 2600, Australia
}
 
\begin{document}

\maketitle

\begin{abstract}

We present and discuss the results of a search for extremely metal-poor stars based on photometry from data release DR1.1 of the SkyMapper imaging survey of the southern sky.   In particular, we outline our photometric selection procedures and describe the low-resolution ($R$ $\approx$ 3000) spectroscopic follow-up observations that are used to provide estimates of effective temperature, surface gravity and metallicity ([Fe/H]) for the candidates.  The selection process is very efficient: of the 2618 candidates with low-resolution spectra that have photometric metallicity estimates less than or equal to --2.0, 41\% have [Fe/H] $\leq$ --2.75 and only $\sim$7\% have [Fe/H] $>$ --2.0 dex.  The most metal-poor candidate in the sample has [Fe/H] $<$ --4.75 and is notably carbon-rich.  Except at the lowest metallicities ([Fe/H] $<$ --4), the stars observed spectroscopically are dominated by a `carbon-normal' population with [C/Fe]$_{1D,LTE}$ $\leq$ +1 dex.  Consideration of the A(C)$_{1D, LTE}$ versus [Fe/H]$_{1D, LTE}$ diagram suggests that the current selection process is strongly biased against stars with A(C)$_{1D, LTE}$ $>$ 7.3 (predominantly CEMP-$s$) while any bias against 
stars with A(C)$_{1D, LTE}$ $<$ 7.3 and [C/Fe]$_{LTE}$ $>$ +1 (predominantly CEMP-no) is not readily quantifiable given the uncertainty in the SkyMapper $v$-band DR1.1 photometry.  We find that the metallicity distribution function of the observed sample has a power-law slope of $\Delta$(Log N)/$\Delta$[Fe/H] = 1.5 $\pm$ 0.1 dex per dex for --4.0 $\leq$ [Fe/H] $\leq$ --2.75, but appears to drop abruptly at [Fe/H] $\approx$ --4.2, in line with previous studies.

\end{abstract}

\begin{keywords}
stars: abundances -- stars: Population II -- Galaxy: halo -- Galaxy: stellar content
\end{keywords}

\section{Introduction} \label{Intro} 

The Big Bang produced hydrogen and helium and small amounts of light elements such as lithium: all other chemical elements have their origin in stellar nucleosynthetic processes.   Such a simple sentence glosses over the enormous amount of physics that underlies the formation and evolution of stars and galaxies, but with observation and theory we can progress towards understanding these processes in detail.  In particular, the stars that formed from the original pristine gas, the so-called Population III stars, are generally thought to have been relatively massive and short-lived \citep[see, for example][and references therein]{Bromm13}.  The
massive stars in this generation ended their lives in supernova explosions and enriched the surrounding gas with their nucleosynthetic products.  No genuine (long-lived and low mass) Population III star has yet been discovered, but we can learn about the properties of these first generation stars through the study of element abundances in the second and subsequent generations, as these contain low-mass long-lived stars that are observable at the present day \citep[see, e.g., the review article of][and references therein]{FN15}.  The rapid rate of star formation at early times in the evolution of the Milky Way with the accompanying rapid increase in stellar metallicity, as well as the Galaxy's subsequent evolution, means that second generation stars are exceedingly rare.  Yet the search for such stars, which are characterized by extremely low abundances of elements such as iron, relative to the solar abundance, has been a quest for decades.

Previous and on-going spectroscopic and photometric surveys for such metal-poor stars include the HK survey \citep{Beers92}, the HES \citep{Christlieb08, Frebel06}, the
ToPos survey \citep{Caffau13}, `Best and Brightest'  \citep{SC14}, the EMBLA survey \citep{LH16}, and more recently the Pristine Survey \citep{Starkenburg17}, the SDSS/BOSS/LAMOST surveys \citep[][and references therein]{Aguado17,Aguado18a}, and the J-PLUS survey, with its southern hemisphere equivalent S-PLUS \citep[][and references therein]{Whitten19}.  However, at the present time only about 30 stars are known with well-established [Fe/H] values that are less than \mbox{--4.0} \citep[e.g.][]{AbduF18}, and of these only five\footnote{A sixth star, SMSS J160540.18--144323.1 first identified in this work, has also been shown to have [Fe/H] $<$ --5; see \citet{TN18}.} have [Fe/H] $<$ --5.0 dex.  These are SMSS J0313--6708 \citep{SK14, MSB15, TN17},
J0023+0307 \citep{Aguado18a, AF19,Aguado19} and J0815+4729 \citep{Aguado18b}, which have upper limits on [Fe/H] of --6.5, --6.3 and --5.8, respectively, HE~1327--2326 with [Fe/H] = --5.5 \citep{AF05, Aoki06} and HE~0107--5240 with [Fe/H] = \mbox{--5.4} \citep{NC02, NC04}.  With such small numbers it is, for example, difficult to distinguish with any statistical rigour between theoretical models that predict an abundance cutoff and a stochastic distribution at lower abundances, from models that predict an on-going smooth decline in the metallicity distribution function.  A larger sample of such stars is clearly required.

As discussed in \citet[][see also \citet{Wolf18}]{SK07}, one of the principal science aims of the SkyMapper Southern Sky 
Survey\footnote{see http://skymapper.anu.edu.au} of
 the southern hemisphere sky is the identification of candidate extremely metal-poor (EMP)\footnote{\citet{BC05} use the terminology `extremely', `ultra' and `hyper' metal-poor to designate stars with [Fe/H] $<$ --3.0, --4.0 and --5.0, respectively.  We have not specifically followed this convention, although our usage of `extremely metal-poor' does generally signify [Fe/H] $\leq$ --3.0.} stars.  This is achieved through the inclusion in the SkyMapper filter set of an intermediate-band $v$ filter ($\lambda_{c}$ = 3825\AA, FWHM = 310\AA) that includes the H- and K-lines of Ca {\sc ii} in the band pass.  At the temperatures corrresponding to F, G, and K spectral types, lower metallicity stars have less blanketing in the $v$-band generating the metallicity sensitivity.  As will be discussed in more detail below, metallicity information can be obtained from SkyMapper photometry via a two-colour diagram in which a metallicity index $m_{i}$, defined as $(v-g)_{0}$ -- 1.5 $(g-i)_{0}$ \citep{SK07}\footnote{Note that the y-axis label in Fig.\ 13 of \citet{SK07} is incorrect.  It should be $(v-g)_{0}$ -- 1.5 $(g-i)_{0}$.}, is plotted against $(g-i)_{0}$, a proxy for effective temperature.

The search for EMP-stars based on SkyMapper photometry was initiated during the commissioning of the telescope.  This `commissioning-era' survey, which is superceded by the current work, led to the discovery of the most iron-poor star known SMSS J031300.36--670839.3 (usually abbreviated to SMSS J0313--6708), which has [Fe/H] $<$ --6.5 (3D, NLTE) \citep{SK14, MSB15, TN17}.  A total of 139 additional Galactic halo EMP 
candidates from the commissioning-era survey were also followed-up at high dispersion to provide detailed abundance information: 44 stars were shown to have  [Fe/H] $\leq$ \mbox{--3.0} and 3 have [Fe/H] $\approx$ --4.0 \citep{HJ15, AFM19}.  Full descriptions of the derived abundances and abundance ratios for these stars are given and discussed in \citet{HJ15} and \citet{AFM19}.  SkyMapper commissioning-era photometry was also employed by \citet{LH16} in a search for EMP stars in the Galactic Bulge.  The most metal-poor object found has [Fe/H] = --3.94 $\pm$ 0.16 based on high dispersion spectroscopic follow-up \citep{LH15}.
In the present work we describe a new search for EMP-stars, based on SkyMapper data release DR1.1\footnote{
http://skymapper.anu.edu.au/data-release/dr1, DOI:~10.4225/41/593620ad5b574}.

The paper is arranged as follows.  In the next section we describe the SkyMapper source photometry and the selection process employed to identify photometric EMP-candidates.  Included in this section is verification of the selection process via globular cluster and known EMP-stars as well as a brief discussion of its limitations.  Section \ref{23m_spect} details the low resolution spectroscopic follow-up of photometric candidates, and the derivation of stellar parameters from the spectra.
Section \ref{MDF} describes the metallicity distribution function resulting from the observed sample of stars and the potential biases that result from the selection and observing processes.  The results are discussed in \S \ref{discuss-sect} while the final section provides a summary and outlines potential future developments in the search process.

\section{Photometric Selection} \label{phot_sect}

We use as the fundamental source for candidates SkyMapper Southern Sky Survey Data Release DR1.1.  As detailed at  http://skymapper.anu.edu.au/data-release/dr1/\#dr1p1\_intro, the release contains photometry in all six SkyMapper bands for objects across an area 
of approximately 17,200 deg$^2$ in the southern hemisphere sky.  In particular, DR1.1 is an update to the original DR1 release (which is no longer publicy available) that provided a significant enhancement to the homogeneity of the photometric calibration across the sky.  The data come from the {\it Shallow Survey} that reaches approximately AB-mag = 18 in the $ugriz$ filters while the limit in the $v$-band is $\sim$0.5 mag brighter \citep{Wolf18}.  These limiting magnitudes are not a concern for the selection of EMP candidates as $g$ $\approx$ 16 is a practical limit for follow-up high dispersion spectroscopy on 8m-class telescopes.

\begin{figure*}
\centering
\includegraphics[angle=0.,width=1.0\textwidth]{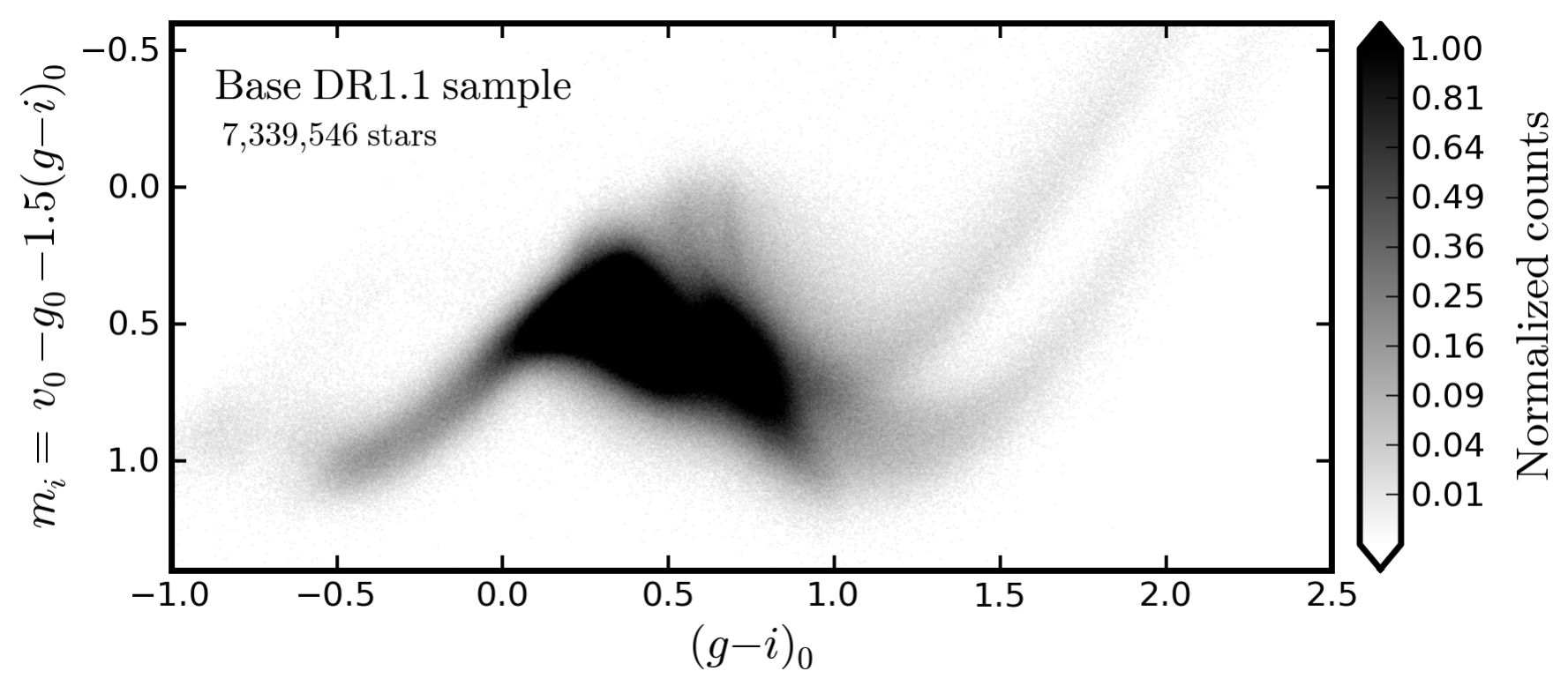}
\caption{The SkyMapper metallicity-sensitive diagram, $m_{i}$ = $(v-g)_{0}$ -- 1.5 $(g-i)_{0}$ versus $(g-i)_{0}$, for all objects satisfying the basic DR1.1 selection criteria.  Metal-poor stars are found at bluer (more negative) values of $m_{i}$ for $\sim$0.3 $\leq$ $(g-i)_{0}$ $\leq$ $\sim$1.0 mag.  The relative (square root) density grey-scale is shown on the right side of the figure.  
}
\label{all_stars_fig}
\end{figure*} 

The initial selection from the DR1.1 database used the following joint criteria \citep[see][for detailed parameter descriptions]{Wolf18}: {\it class-star} $>$ 0.9 to ensure that the object is stellar in nature; $flags$ $\leq$ 3 so that there are no apparent issues with the SExtractor photometry; {\it nch\_max} = 1 so that there is a single source for each filter\footnote{Given the seeing at Siding Spring Observatory \citep[e.g. the median seeing for the $g$-band DR1 images is 2$\farcs$6,][]{Wolf18} this turned out to be an important criterion -- not infrequently a poor-seeing $v$-image is classified as two stars which ultimately results in photometry that generates a spurious EMP candidate.  Implementation of this criterion significantly reduced the contamination of the 2.3m spectroscopy by relatively metal-rich stars.}; {\it ngood\_min} $\geq$ 1 $AND$ {\it v\_ngood} $\geq$ 2 to ensure at least one measurement in all of $ugriz$ and at least two measurements in the $v$-band given its importance; {\it g$_{psf}$} $\leq$ 16.1 so that candidates can be followed-up at high dispersion on 8m-class telescopes in reasonable integration times; {\it e\_g\_psf} AND {\it e\_i\_psf} $<$ 0.03 and {\it e\_v\_psf} $<$ 0.05 mag as a compromise between photometric precision and number of candidates selected; {\it ebmv\_sfd} $<$ 0.25 mag to avoid large reddening corrections to the photometry and to avoid areas of the sky where the photometry is less well calibrated and frequently affected by image crowding; and, {\it prox} $>$ 7.5 AND {\it twomass\_dist2} $>$ 7.5 so that there is no other DR1.1 or additional 2MASS source within 7$\farcs$5 of the target.   

Application of these criteria to the DR1.1 database then resulted in over 7 million candidates\footnote{Such a sample cannot be downloaded in one-step given the current million-row limit on queries.  In practice the selection was run a number of times for specific intervals in RA.}.  The metallicity-index $m_{i}$ = $(v-g)_{0}$ -- 1.5 $(g-i)_{0}$ versus $(g-i)_{0}$ colour-colour diagram for these stars is shown in Fig.\ \ref{all_stars_fig}.  In calculating the reddening corrected magnitudes we have used the rescaled E$(B-V$)$_{SFD}$ colour excesses, calculated as described in \citet{Wolf18}, together with the reddening coefficients given in that paper.  Specifically, we employ $A_v$, $A_g$ and $A_i$ values of 4.026, 2.986 and 1.588, respectively \citep{Wolf18}.

Inspection of the figure shows that there is a cloud of stars that extends to bluer (i.e. more negative) values of $m_{i}$ at a given $(g-i)_{0}$, particularly for the approximate colour range 0.3 $\leq$ $(g-i)_{0}$ $\leq$ 1.0.  This is where we expect the index to be sensitive to metallicity \citep{SK07}, and thus this region is where the EMP-candidates are expected to occur.  Note that beyond $(g-i)_{0}$ $\approx$ 1.2 a bifurcation is visible in this diagram as the index becomes primarily gravity rather than metallicity sensitive; the upper sequence is for cool dwarfs and the lower for cool giants both at approximately solar metallicity.  We now turn to verifying and calibrating the metallicity-index diagram in order to be able to refine the selection to generate a tractable sample.

\subsection{Verification}

Since the SkyMapper photometry, particularly in regard to the $v$ magnitudes, is effectively a new photometric system, it is important to verify the relation between location in the metallicity-sensitive diagram and abundance, which we represent by [Fe/H].  We do this particularly for metal-poor stars since these are our focus.  Other studies, such as \citet{Luca18}, have investigated the situation at higher metallicities. 

\subsubsection{Metal-poor Globular Cluster Red Giants}

\begin{figure}
\centering
\includegraphics[angle=0.,width=0.48\textwidth]{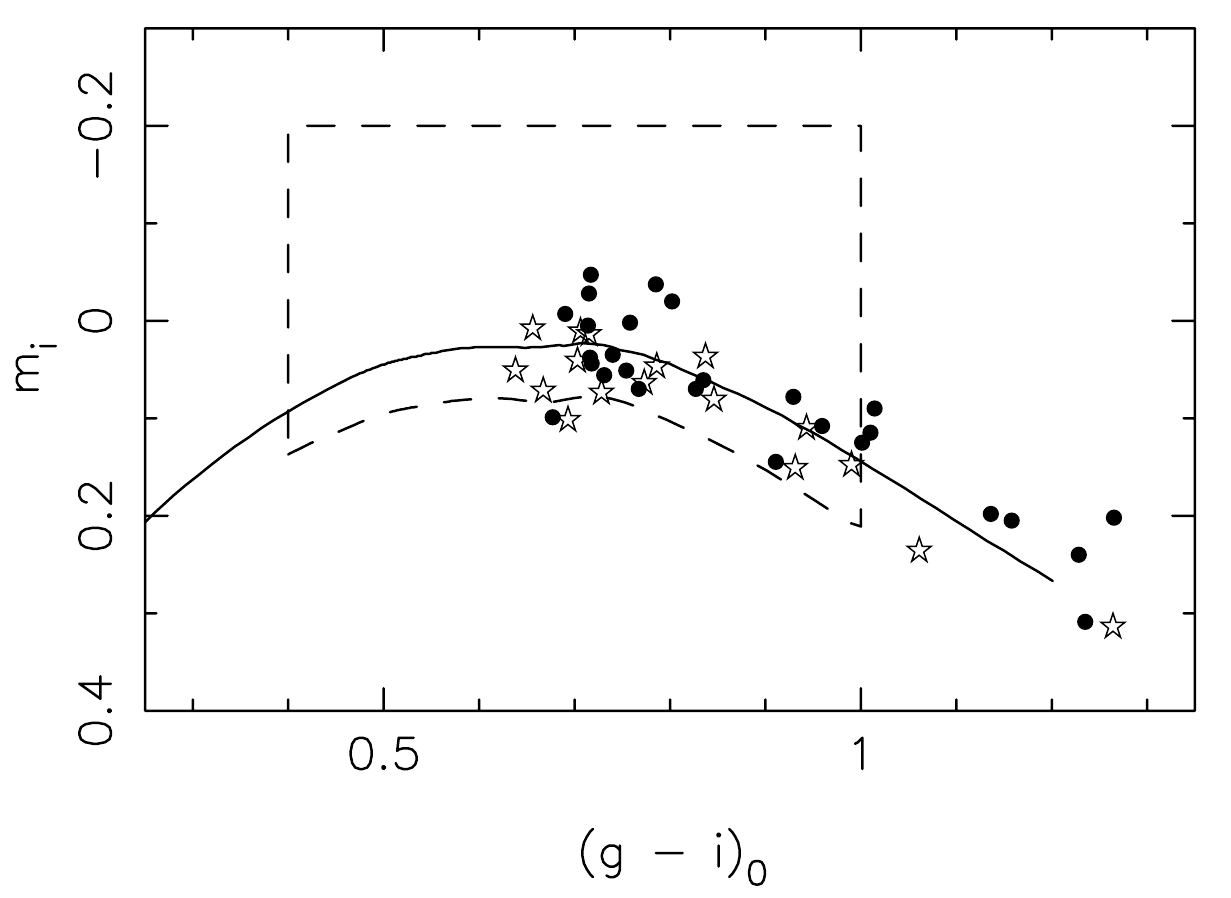}
\caption{SkyMapper metallicity-sensitive diagram for probable red-giant members of the metal-poor globular clusters NGC~4590 (filled circles) and NGC~7099 (open star symbols) compared with a 12.5 Gyr, [Fe/H] = --2.25, [$\alpha$/Fe] = +0.4 Dartmouth isochrone.  The dashed outline is the final adopted selection window as discussed in \S \ref{ASW}.}
\label{cmd_fig_4590}
\end{figure}

NGC~4590 (M68) and NGC~7099 (M30) are among the most metal-poor globular clusters in the Galaxy: the most recent 
on-line\footnote{http://physwww.physics.mcmaster.ca/$\sim$harris/mwgc.dat}
version of the catalogue of \citet{H96} lists metallicities of [Fe/H] = --2.23 and --2.27, respectively.  Given the similar metallicities the natural expectation is that the cluster red giants should define the location of an iso-metallicity locus in the metallicity-sensitive diagram.  To investigate this we have cross-matched red giant branch (RGB) stars for these clusters selected from the photometric  standards database maintained by Dr.\ Peter Stetson\footnote{http://www3.cadc-ccda.hia-iha.nrc-cnrc.gc.ca/community/STETSON/standards/}  with the SkyMapper DR1.1 database, employing essentially the same selection criteria as above.  The specific stars used are generally known cluster members from AAT/AAOmega spectroscopy \citep[e.g.][]{DaC16} that are brighter than the cluster horizontal branch.  The location of these globular cluster red giants in the metallicity-sensitive diagram is shown in Fig.\ \ref{cmd_fig_4590}.  Also shown in the figure is the location of a theoretical isochrone for an age of 12.5 Gyr, abundance
[Fe/H] = --2.25 and [$\alpha$/Fe] = +0.4 from the Dartmouth Stellar Evolution Database isochrone set \citep{AD08}.

The agreement between the observations and the theoretical isochrone is good: for 0.6 $\leq$ $(g-i)_{0}$ $\leq$ 0.8 where the isochrone is approximately constant at $m_{i}$ = 0.029 mag, the mean of the globular cluster stars is $m_{i}$ = 0.033 with a standard deviation $\sigma$ = 0.040 mag.  The observed scatter in the globular cluster $m_{i}$ values is large but there is no obvious indication of a systematic offset between the observations and the theoretical isochrone.  The scatter is presumably driven by uncertainties in the $v$ magnitudes as the corresponding $i_{0}, (g-i)_{0}$ combined colour-magnitude diagram (generated using an appropriate relative distance modulus offset) shows only a small scatter in $(g-i)_{0}$: $\sigma (g-i)_{0}$ = 0.021 mag about the mean RGB locus.

\subsubsection{Known EMP-stars} \label{knownEMP}

\begin{figure*}
\centering
\includegraphics[angle=0.,width=0.75\textwidth]{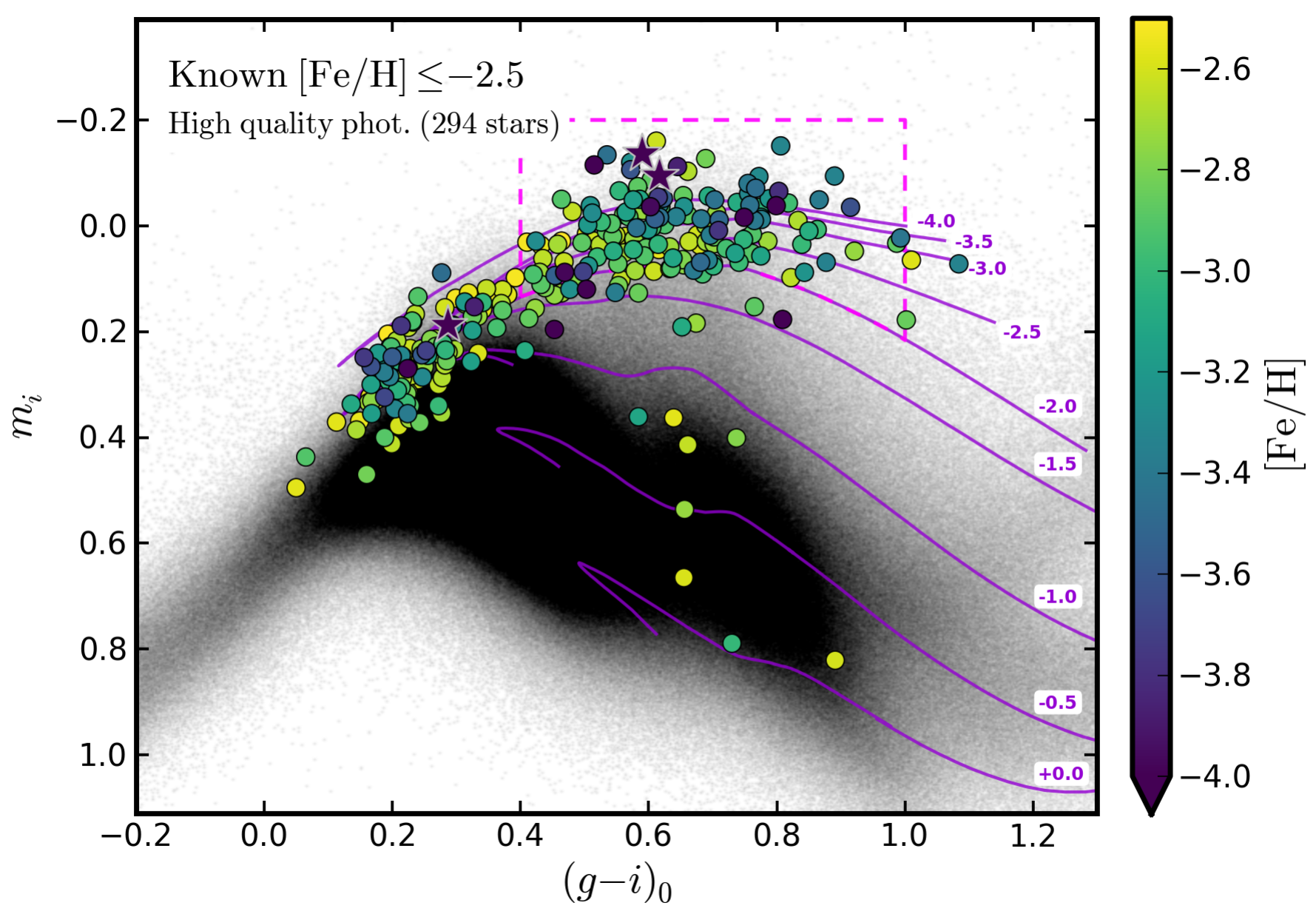}
\caption{SkyMapper metallicity-sensitive diagram for all objects satisfying the basic selection criteria.  Overplotted as coloured circles are the locations in this diagram of known EMP-stars with [Fe/H] $\leq$ --2.5, determined from high dispersion spectroscopic studies, and with photometry in SkyMapper data release DR1.1 meeting our criteria.  The colour-bar giving the corresponding high dispersion spectroscopy derived metallicities is on the right side of the figure.  Shown also are the location of isochrones from the Dartmouth Stellar Evolution Database for the metallicities, in order of decreasing metallicity index,  [Fe/H] = 0.0, --0.5, --1.0, --1.5, --2.0, --2.5, --3.0, --3.5, and --4.0.  The isochrones are for an age of 12.5 Gyr and have [$\alpha$/Fe] = 0.0 for [Fe/H] = 0.0 and --0.5, [$\alpha$/Fe] = +0.2 for [Fe/H] = --1.0 and [$\alpha$/Fe] = +0.4 for all other metallicities. They are labeled with the metallicity.  Note that the lower main sequence portions of the isochrones are not plotted.  The hyper metal-poor stars SMSS J031300.36--670839.3, HE~0107--5240 and HE~1327--2326 lie at ($(g-i)_{0}, m_{i}$) = (0.62, --0.09), (0.59, --0.135) and (0.29, 0.19), respectively, and are plotted as 5-point star symbols with the appropriate colour. The final adopted selection window is shown by the magenta dashed lines.}
\label{cmd_knownEMP}
\end{figure*}

An alternative approach is to investigate the location of known EMP-stars in the metallicity-sensitive diagram.  To perform this task we have compiled from the literature a list of stars with [Fe/H] $\leq$ --2.5, as determined from high-dispersion spectroscopy.
The list was then cross-matched with the SkyMapper DR1.1 database.  In carrying-out the match-up we have used tighter limits on the SkyMapper photometry in order to reduce the effects of photometric uncertainty on the location of the stars in the metallicity-sensitive diagram.  Specifically, we required {\it e\_g\_psf} AND {\it e\_i\_psf} $\leq$ 0.02,  {\it e\_v\_psf} $\leq$ 0.03, and E$(B-V$)$_{SFD}$ $<$ 0.1 mag,  and this resulted in a sample of 294 stars.  The majority of the stars are drawn from \citet{DY13}, \citet{HJ15}, \citet{Barklem2005}, \citet{Roed14} and \citet{Cohen13}; the effective temperatures, gravities and metallicities from the latter two sources have been adjusted to ensure consistency with the other sources.  The location of the known-abundance stars in the metallicity-sensitive diagram is shown in Fig.\ \ref{cmd_knownEMP}.  We note that the most iron-poor star known, star SMSS J031300.36--670839.3, is the dark coloured 5-point star symbol at ($(g-i)_{0}, m_{i}$) = (0.62, --0.09) in the figure, while the star HE~0107--5240, which has [Fe/H] $\approx$ --5.4, is the dark coloured 5-point star symbol at (0.59, --0.135).  The sub-giant star HE~1327--2326 ([Fe/H] $\approx$ --5.7) is located at (0.29, 0.19), and is also plotted as a dark coloured 5-point star symbol
 in Fig.\ \ref{cmd_knownEMP}.  In general, the majority of the stars plotted are consistent with expectations in that they lie at notably bluer $m_{i}$ values at a given $(g-i)_{0}$ than the bulk of the population.  However, this is clearly not the case for a small number of known-EMP stars that fall well below their `expected' location in the metallicity-sensitive diagram.  As will be discussed in more detail in \S \ref{Cabund_sect}, generally these stars are strongly enhanced in carbon.

Fig.\ \ref{cmd_knownEMP} also shows isochrones from the Dartmouth Stellar Evolution Database isochrone set \citep{AD08} for metallicities ranging from solar to [Fe/H] = --4.0 dex\footnote{The Dartmouth Stellar Evolution Database contains isochrones only to a metallicity [M/H] = --2.5 dex.  The isochrones for lower [M/H] values were calculated specifically for us by Dr.\ Aaron Dotter, but are otherwise equivalent to those in the database.}.  
In each case the adopted age is 12.5 Gyr and the
[$\alpha$/Fe] values assumed are 0.0 for [Fe/H] = 0.0 and --0.5, +0.2 for [Fe/H] = --1.0 and [$\alpha$/Fe] = +0.4 for all other metallicities.  Note that only a portion of the full isochrone is shown in each case: the section from just below the main sequence turnoff to the tip of the RGB\@.  We have chosen not to plot the lower main sequence portion of the isochrones as, given the relatively bright apparent magnitude cutoff at $g$ $\approx$ 16, extremely metal-poor cool dwarfs are unlikely to occur in the survey in any great number given their intrinsically low luminosity.
 
In order to test the agreement of the location of the isochrones and the known EMP-star photometry, we first selected from the sample of 294 stars those with 0.4 $\leq$ $(g-i)_{0}$ $\leq$ 1.0 for which [C/Fe] was measured and had a value in the range --1.0 $<$ [C/Fe] $<$ +0.7 dex (i.e., we excluded carbon-enhanced metal-poor (CEMP) stars).  We further required that the $T_{\rm eff}$ and log $g$ values were consistent with a subgiant or red giant branch classification in order to remove any dwarfs and any post-AGB stars.  The stars were then split into groups of $\pm$0.25 dex in metallicity centered at [Fe/H] = --2.5, \mbox{--3.0}, \mbox{--3.5} and --4.0, respectively. 
We then computed for each group the mean value of the difference between the $m_{i}$ index for the star and that of the corresponding isochrone at the star's $(g-i)_{0}$ colour, employing mild sigma-clipping ($\pm$2$\sigma$) in the calculation of the means.  Unlike the case for the metal-poor globular clusters, the offsets were significantly different from zero with values of 0.019 ($\sigma$ = 0.027, N=41), 0.037 (0.040, 59), 0.033 (0.045, 30), and 0.028 (0.048, 7), respectively, in the sense that the stars lie at larger values of $m_{i}$.  

Given the consistency of the offsets, we decided to shift the isochrones redder in $m_{i}$ by 0.027 mag, the overall mean offset value.  Such an offset is not unexpected given the uncertainty in the overall zero point of the DR1.1 $v$ magnitudes \citep{Wolf18,Luca18}.  The isochrones shown in Fig.\ \ref{cmd_knownEMP} have had this offset applied, and for simplicity we have assumed that it also applies to the isochrones with [Fe/H] = --2.0 and higher.  Regarding the metal-poor globular cluster red giants, application of the offset means that there is now a systematic difference (--0.023 mag relative to the shifted isochrones), but that difference remains within a 1$\sigma$ uncertainty given the spread in the globular cluster star $m_{i}$ values.  The difference in inferred isochrone offsets between the globular cluster stars and the known abundance stars may well be an indicator of field-to-field variations in the zero point of the DR1.1 $v$ photometry, in which case the offset derived from the known metallicity stars is the appropriate one to apply as it is based on stars that are scattered across many SkyMapper fields. 

These same groups of known metallicity stars, which we note have been selected to lie on the subgiant or red giant branches, have precise photometry and low reddening, and which exclude CEMP stars, also demonstrate that the $m_{i}$ index retains
metallicity sensitivity at low metallicities.  In particular, the mean offsets in $m_{i}$ from the adjusted [M/H] = --2.0 isochrone are --0.054,  --0.070, --0.102 and \mbox{--0.138} mag for the groups centered on metallicities of \mbox{--2.5}, \mbox{--3.0}, --3.5 and --4.0, respectively.  
Given that the standard errors for these mean values are less than 0.01 mag (the standard deviations and numbers are given above), the values demonstrate conclusively that, for precise photometry and excluding CEMP stars, the $m_{i}$ index does retain metallicity sensitivity to low abundances.

For completeness we note that if the isochrone offset exercise is repeated for the bluer stars (i.e., those with $(g-i)_{0}$ $\leq$ 0.4), which are predominantly lower subgiant branch and main sequence turnoff stars, then the computed offset is 2-3$\times$ larger though less well determined because of the smaller number of stars.  This effect is evident in Fig.\ \ref{cmd_knownEMP} where the isochrones tend to lie above the stars in the vicinity of the main sequence turnoff.

\subsection{The Adopted Selection Window} \label{ASW}

We now consider how to use the information presented in Fig.\ \ref{cmd_knownEMP} to define a photometric sample of EMP-candidates that is practical in size, a necessity driven by the uncertainties in the $v$ magnitudes which are sufficiently large that we cannot simply take the stars with the lowest $m_{i}$ values directly to a 8m-class telescope for high dispersion spectroscopic follow-up.  An intermediate winnowing step is necessary, and that requires a tractable sample to work from.  We proceed as follows: first, the location of the --2.5 and lower metallicity isochrones in Fig.\ \ref{cmd_knownEMP} suggest that the RGB at low metallicities does not go substantially redder than $(g-i)_{0}$ $\approx$ 1.0 to 1.2 mag.  Further, stars near the RGB-tip have the lowest temperatures and gravities and are therefore somewhat more complicated as regards atmospheric abundance analyses.  The luminosity function on the RGB is also declining steeply as the RGB-tip is approached.  Consequently, given these factors, we have chosen to limit our selection window to stars with $(g-i)_{0}$ $\leq$ 1.0 ($T_{\rm eff}$ $\approx$ 4550 K).

In a similar fashion, inspection of Fig.\ \ref{cmd_knownEMP} also reveals that at bluer colours, $(g-i)_{0}$ $\leq$ $\sim$0.4 mag, the separation between the known EMP stars and the bulk of the stars (here presumably relatively young approximately solar metallicity disk dwarfs) is much smaller than it is for redder colours.  EMP-stars are inherently rare so that any investigation of stars in this $(g-i)_{0}$ colour regime, even at blue ${m_{i}}$ colours, is likely to be significantly contaminated by non-EMP objects.  For that reason we have chosen to place the blue colour selection limit at $(g-i)_{0}$ = 0.4 mag ($T_{\rm eff}$ $\approx$ 5750 K).  We recognise that in adopting this colour cut we are selecting against EMP-stars in the vicinity of the main sequence turnoff. 

As regards limits on the ${m_{i}}$ values, we have chosen to select only stars with ${m_{i}}$ $>$ --0.2 mag.  As Fig.\ \ref{cmd_knownEMP} shows, the isochrones become increasingly closer together as the abundance drops, and the known EMP-stars are within this limit.  Limited low-resolution spectroscopic follow-up (see \S \ref{23m_spect}) of objects with 
${m_{i}}$ $\leq$ --0.2 shows that they are frequently young stars with Ca {\sc ii} H+K emission or extragalactic objects such as QSOs and AGNs.  

\begin{figure*}
\centering
\includegraphics[angle=0.,width=0.95\textwidth]{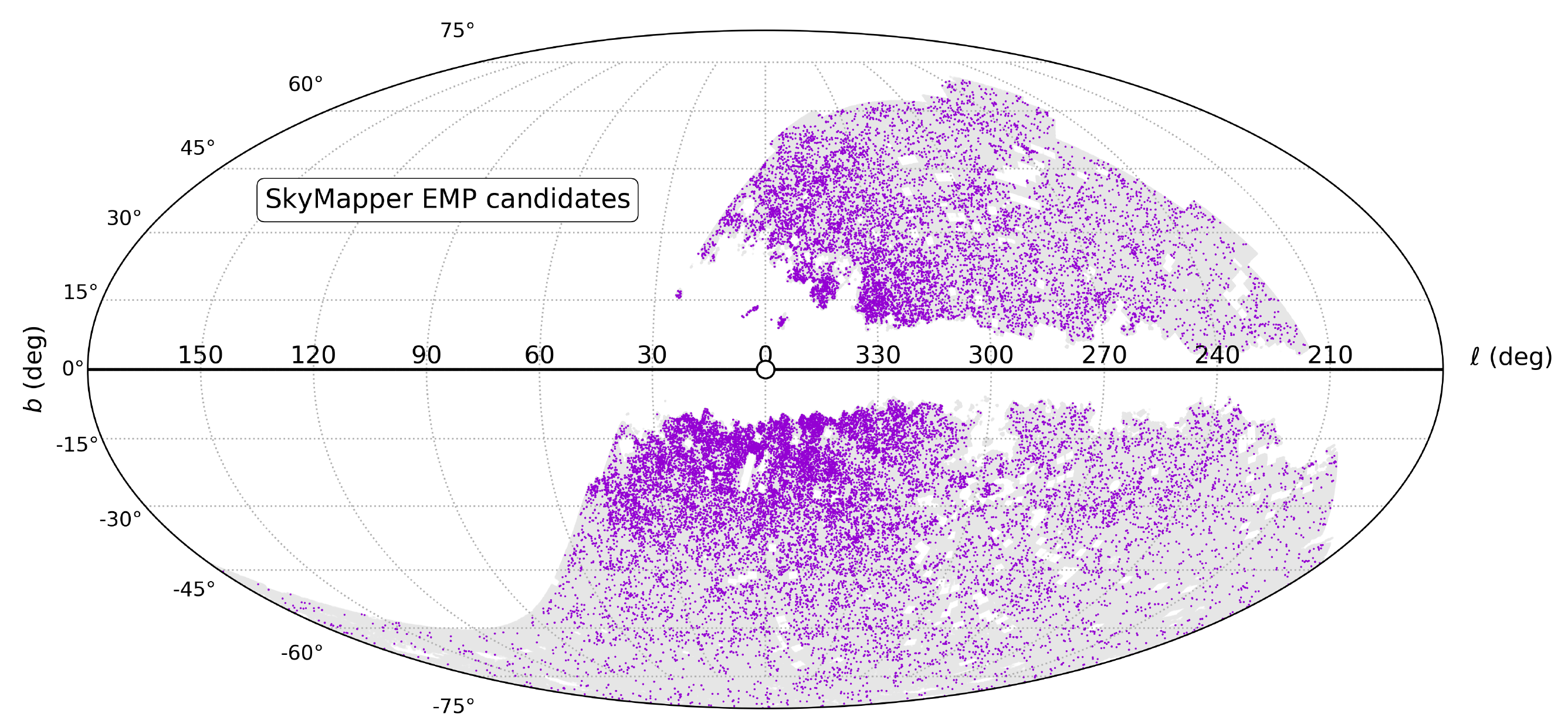}
\caption{A Mollweide projection of the southern sky in Galactic coordinates on which the SkyMapper DR1.1 EMP photometric candidates are plotted as small purple dots on top of the complete photometric selection (grey background).  White areas indicate regions where SkyMapper DR1.1 data are lacking or incomplete. 
The EMP candidates are clearly concentrated towards the Galactic Centre as would be expected for a population dominated by halo stars.  }
\label{allsky_fig}
\end{figure*} 

The most crucial choice for the EMP-selection window is the placing of the red (more positive) boundary for the ${m_{i}}$ values.  As is evident in Fig.\ \ref{cmd_knownEMP}, at fixed $(g-i)_{0}$ the number of potential candidates increases extremely rapidly with increasing 
${m_{i}}$.  A relatively high value for the ${m_{i}}$ limit would therefore yield a large number of candidates, but would likely not miss any genuine EMP-stars assuming they could be found within the much more numerous non-EMP stars.  On the other hand, a bluer (lower) value for the ${m_{i}}$ cut  produces a smaller sample but possibly misses some genuine EMP stars whose 
${m_{i}}$ are erronously high.  We decided to compromise by adopting the (adjusted) location of the --2.0 isochrone as the red
boundary in ${m_{i}}$ for the selection window.  The final adopted selection window is outlined by the dashed boundary shown in Figs.\ \ref{cmd_fig_4590} and \ref{cmd_knownEMP}.

It is worth emphasizing that the selection process, and the consequent contamination of the selected sample by more metal-rich objects, is essentially driven by uncertainties in the $v$-band photometry.  Replacing $(g-i)_{0}$ with other Skymapper colours, or non-Skymapper photometry such as the {\it Gaia} DR2 $(G_{BP}  - G_{RP})_{0}$ colour, does not make any appreciable difference to the sample selection.  The $v$-band errors result from both intrinsic measurement error from the 20 sec exposures and the uncertainty in the uniformity of the calibration of the zero point across the sky.  \cite{Luca18} has shown, for example, that there are systematic zero points trends with Galactic latitude for both $u$ and $v$ in the SkyMapper DR1.1 photometry.

Applying the adopted selection window to our SkyMapper DR1.1 sample then results in a total sample of 26,600 
stars with $g$ $\leq$ 16.1 that are predicted on the basis of the photometry to have [Fe/H] $<$ --2.0; underlying uncertainties in the photometry mean that attempting to calculate specific metallicities for specific stars from the photometry is not appropriate or meaningful.  Since we are seeking new EMP candidates, we then excluded from the sample 232 stars that have existing [Fe/H] estimates, whether from high or low dispersion spectroscopy.  We also excluded the 191 candidates that lie within the tidal radius of a Galactic globular cluster, using the information given in the current on-line version of the \citet{H96} catalogue.  This latter approach may exclude some genuine EMP-stars but it avoids accidental rediscovery of metal-poor globular cluster members.  After this process 26,237 stars remain of which 15925 have 15.0 $<$ $g$ $\leq$ 16.1; 9,442 have 13.0 $<$ $g$ $\leq$ 15.0 and 870 are brighter than $g$ = 13.0 mag.  These data sets then form the basis of the low-resolution spectroscopic program that endeavours to separate the EMP-star `wheat' from the contaminating `chaff'.  

Because of the restriction on the allowed reddening of the candidates, the selected stars are expected to be predominantly halo, as distinct from disk, red giants and this expectation is confirmed by the distribution of the photometric EMP-candidates on the sky.  As shown in Fig.\ \ref{allsky_fig}, the on-sky distribution of the candidates shows a clear concentration in the direction of the Galactic Centre, consistent with a population dominated by halo stars.  Fig.\ \ref{allsky_fig} also shows some potential structure or grouping of EMP candidates on smaller scales, but we are reluctant to consider any such sub-structure as real until the large-spatial-scale uncertainties in the $v$-band photometry are minimized.

\section{Low Resolution Spectroscopic Follow-up} \label{23m_spect}

Commencing in 2016 February we have observed SkyMapper photometric EMP-candidates with the ANU 2.3m telescope at Siding Spring Observatory on a regular basis.  The program has been allocated approximately 6 nights per month and useful data has been obtained for $\sim$60\% of the time.  The program is on-going and the results presented here include data from observing runs up to and including 2018 November.  

The 2.3m observing program commenced with candidates selected from an initial internal pre-release version of the SkyMapper Early Data Release (EDR) and subsequently from the EDR itself (DOI: 10.4225/41/572FF2C5EBD30) when it was made accessible to the Australian Astronomy community on 2016 May 9.  The EDR covered about one-third of the southern hemisphere sky.  The source of candidates then moved to Data Release 1 (DR1) when it became available on 2017 June 6 \citep{Wolf18}.  This release was replaced by DR1.1 on 2017 Dec 13.  DR1.1 is world-wide accessible.  

Since the release of DR1.1, it has been the source of the photometric EMP-candidates, chosen using the selection criteria and the selection window discussed above.  However, because of variations in the photometry databases and in the basis for selecting candidates, a number of stars have been observed at the 2.3m prior to the release of DR1.1 that either fail the adopted selection criteria, or which fall outside the selection window.  For the present we will ignore these stars and concentrate on those whose DR1.1 characteristics are such that they meet the selection criteria and which fall within the adopted selection window.  As of 2018 November there are 2618 individual EMP-candidates (from the total of 26,237) that meet these criteria and which have at least one 2.3m spectroscopic observation.   A further 937 stars, that do not meet the final photometric selection criteria, also have at least one 2.3m spectroscopic observation.

The 2.3m observations are conducted with the WiFeS integral field spectrograph \citep{MD10} using the B3000 and R3000 gratings to yield resolution $R$ $\approx$ 3000 spectra.  The blue spectra cover the wavelength interval $\lambda$3400--5800\AA\/ and exposure times are set to yield a S/N per pix of $\sim$20 at the H and K lines of Ca\,{\sc ii}.  Because WiFeS is an integral field spectrograph useful spectra can still be obtained in poor-seeing conditions.   The observations were carried out by a number of
different observers.  No specific priority was given to any of the candidates in the selection window, but as discussed in 
detail in \S \ref{Obs_Samp}, in practice the observers tended to prefer candidates with more negative ${m_{i}}$ values for observation, in the expectation that such an approach would enhance the chances of finding extremely metal-poor stars.  

\subsection{Spectrophotometric fits}

The raw observed spectra are extracted, sky-subtracted, wavelength-calibrated and most importantly, (relative) flux-calibrated via observations of a number of known flux standards each night.   Specifically, all stars are observed with the atmospheric
dispersion direction parallel to the IFU slits to minimize flux loss, and the majority of the observations were carried out at relatively low airmass (X $\leq$ 1.4).  The flux standards used are a subset of those in the {\it Hubble Space Telescope} CALSPEC database 
\citep[see][]{Bohlin+14} which have well-established flux distributions at optical wavlengths.  Typically at least two observations of four or more flux standards are observed each night, and we find that the calibrations from the different standards agree well.    
The (relative) absolute flux calibration used is the mean from all the standards observed each night, and we find that the calibration is stable from night-to-night and from month-to-month.  In particular, the slope from $\lambda$4000\AA\/ to 5800\AA\/ in the fluxed spectra, which is sensitive to both temperature and gravity for an assumed reddening,  is very well defined. 

The flux calibrated spectra are then compared with a grid of MARCS 1D model atmosphere fluxes \citep{Gust08} using the {\it fitter} code, described in \citet{JEN13a}, and the best-fit determined.  Specifically, the model fluxes are convolved to a resolution equivalent to the observations and normalized by 
the mean flux between $\lambda$4500 and 5500\AA.  The observed spectrum is shifted to rest wavelength using the velocity determined from a cross-correlation with a standard set of model spectra, and is then also normalized, again by the mean flux in the $\lambda$4500 and 5500\AA\/ region.  As outlined in \citet{JEN13a}, in the first pass the model ($\alpha$-enhanced for [M/H] $\leq$ --1.5) fluxes are interpolated to produce a new grid with spacings of 100 K in $T_{\rm eff}$, 0.5 in log~$g$ and 0.25 dex in [M/H].  Each model spectrum is then compared with the observed spectrum via a simple $\chi ^2$ calculation with the minimum used to determine the parameters of the best-fitting model.  
 In making these initial fits no assumption is made regarding the effective temperature or the metallicity of the star.

A new grid centered on these parameters is then generated with a finer spacing of 25 K in $T_{\rm eff}$ and 0.125 in log~$g$.  The $\chi ^2$ calculation process is then repeated to select the final best-fitting parameters.  An example of the {\it fitter} output is shown in Fig.\ 5 of \citet{JEN13a}.  In computing the best-fits the effect of reddening is also considered, with the initial choice for the adopted reddening being that from \citet{SFD98} modified (reduced) as described in \citet{Wolf18}.  By deliberate choice 
(see \S \ref{phot_sect}), 
E$(B-V)_{SFD}$ $<$ 0.25, so that E$(B-V)_{adopted}$ $<$ 0.18 mag.  Generally the fit with the adopted reddening is satisfactory; however, the residuals at the Balmer lines are inspected and used to judge whether the adopted reddening needs to be modified, in which case the fitting process is reperated.  Overall, uncertainty in the appropriate reddening value is a larger contributor to the uncertainty in the derived temperatures compared to any uncertainty arising from the flux calibration.

\begin{figure}
\centering
\includegraphics[angle=0.,width=0.48\textwidth]{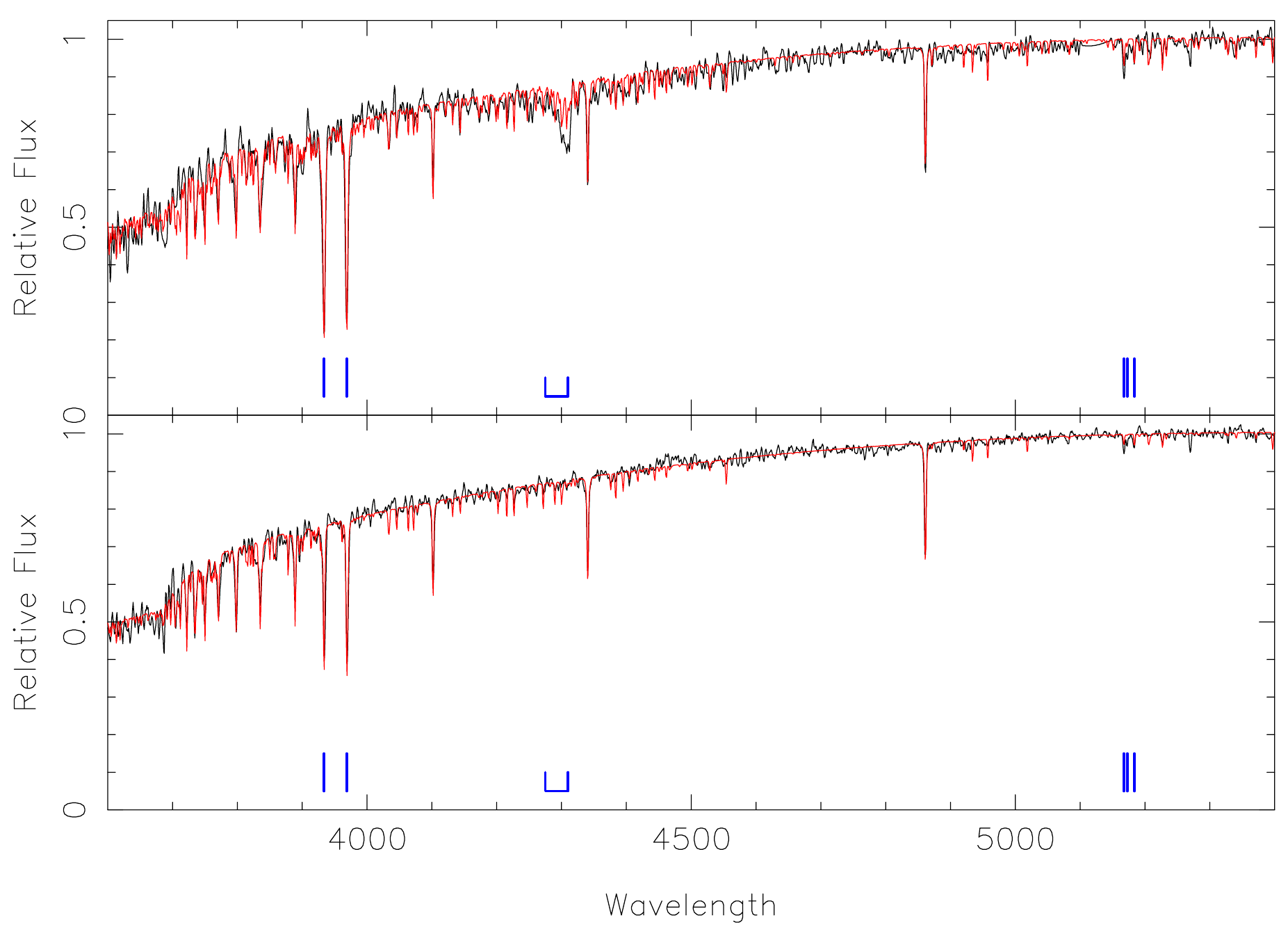}
\caption{Example spectra (black lines) from the 2.3m WiFeS observations of EMP-candidates.  The relative flux values have been normalized to unity at $\lambda$ $\approx$ 5500\AA.  The stars shown are SMSS J121615.75--281249.1 (upper panel, $g$=15.59) and SMSS J182845.79--354022.8 (lower panel, $g$=14.36).  The spectrophotometric fitting process yields 4925/1.625/--3.25 and 4925/1.5/--4.0 for $T_{\rm eff}$, log $g$ and [Fe/H], respectively, and the best-fit model spectra are over-plotted as red lines.  In each panel the wavelengths of the Ca\,{\sc ii} H and K lines ($\lambda$3933, 3986\AA) and the Mg\,{\sc i} b lines ($\lambda$5167, 5172, 5183\AA), are marked, as is the wavelength of the G-band (CH, $\sim$4300\AA).  The fit in the upper panel uses E$(B-V)$ = 0.0, while that in the lower panel employs E$(B-V)$ = 0.1 mag. }
\label{spec_fig}
\end{figure}

Overall, the best-fit temperature and gravity are generally well determined.  Only in the small number of cases where the S/N of the spectra is low, particularly in the vicinity of the Balmer jump so that the gravity is not well constrained, do we move the $T_{\rm eff}$, log $g$ estimate along the valley of the minimum $\chi ^2$ to force agreement with a metal-poor giant branch isochrone.  In this process the $T_{\rm eff}$ estimate changes by no more than 25--50 K but the log $g$ change can be significant. 
The metallicity, quantized at the 0.25 dex level, then follows by maximizing agreement between the observed and the model spectrum with the best-fit temperature and gravity, using the strengths of metal-lines particularly, in the metal-poor regime, the Ca\,{\sc ii} H and K and Mg\,{\sc i} b features.  However,  because of the known diversity in the [C/Fe] ratio among metal-poor stars, the region of the G-band of CH is generally not used in determining the best-fit metallicity value.  Examples of observed spectra and the
corresponding spectrophotometric fits are shown Fig.\ \ref{spec_fig}.

\begin{figure}
\centering
\includegraphics[angle=0.,width=0.48\textwidth]{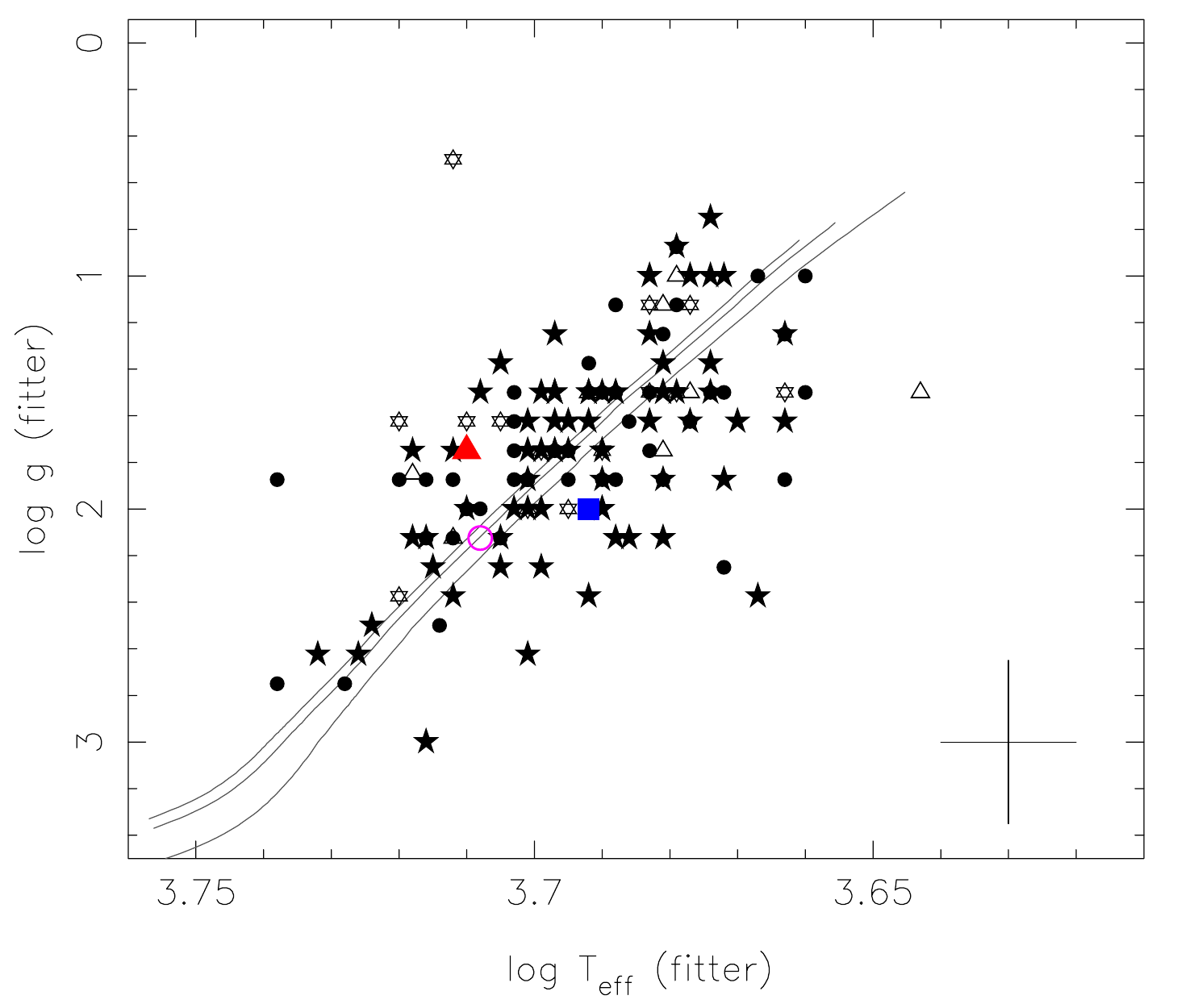}
\caption{The relation between the log $T_{\rm eff}$ and log $g$ values, derived from the spectrophotometric fits, for the 142 stars with [Fe/H]$_{fitter}$ $<$ --3.0 dex.  Symbol code is the following: [Fe/H]$_{fitter}$ = --4.75, blue filled square (SMSS J160540.18--144323.1, see Table \ref{tab:data}); --4.5, red filled triangle (SMSS J054650.97--471407.8); --4.25, magenta open circle
(SMSS J205001.91--661329.7); --4.0, open triangles; --3.75, 6-point star symbols;  --3.5, filled circles, and --3.25, filled stars.  Shown also are the relations for  metal-poor isochrones with [M/H] = --4.0, --3.5, --2.5 and  [$\alpha$/Fe] = +0.4 for an age of 12.5 Gyr. 
$\pm$1$\sigma$ error bars for both quantities are shown in the lower right of the panel.}
\label{logTeff-logg}
\end{figure}

We now concentrate primarily on the 142 stars that meet the photometric selection criteria,  that have [Fe/H]$_{fitter}$ $<$ --3.0  and for which, as far as we are aware, there are no published abundance studies based on high-dispersion spectra.    In Fig.\ \ref{logTeff-logg} we plot the {\it fitter}
log~$g$ values against log~$T_{\rm eff}$.  As expected for a sample that is presumably dominated by halo red giants, the temperature-gravity relation is consistent with that for the red giant branch of an old metal-poor isochrone: the figure shows the isochrones for  [M/H] = --4.0, --3.5 and --2.5 with [$\alpha$/Fe] = +0.4 and age = 12.5 Gyr (see \S \ref{knownEMP}).    Given the errors, $\pm$0.01 in log~$T_{\rm eff}$ and $\pm$0.35 in log~$g$, the agreement with the isochrones is satisfactory, and there is only one definite outlier in this plot -- the six-point star symbol at (log~$T_{\rm eff}$, log $g$) = (3.712, 0.5), which is SMSS J100231.91--461027.5.

We have subsequently recognised this object as an unusual star: it is the variable
ASASSN-V J100232.04--461027.9 \citep{Jay18} that has a well-determined Cepheid-like light curve with a period of 26.597d and $V$ amplitude of 0.54 mag.  The star has a very strong mid-IR excess with WISE  W1, W2, W3 and W4 magnitudes of 9.89, 8.73, 4.31 and 2.03, respectively, indicating large amounts of circumstellar dust.  The star is most likely a post-AGB star in which depletion onto dust grains is responsible for the apparently low photospheric abundances, particularly for the elements with high condensation temperatures \citep[e.g.][]{ML92,Aoki17}.  Given the likely post-AGB nature of the star, the $T_{\rm eff}$ and log $g$ 
values estimated from the 2.3m low-resolution spectrum are likely valid, i.e., consistent with a post-AGB evolutionary track.

\begin{figure}
\centering
\includegraphics[angle=0.,width=0.48\textwidth]{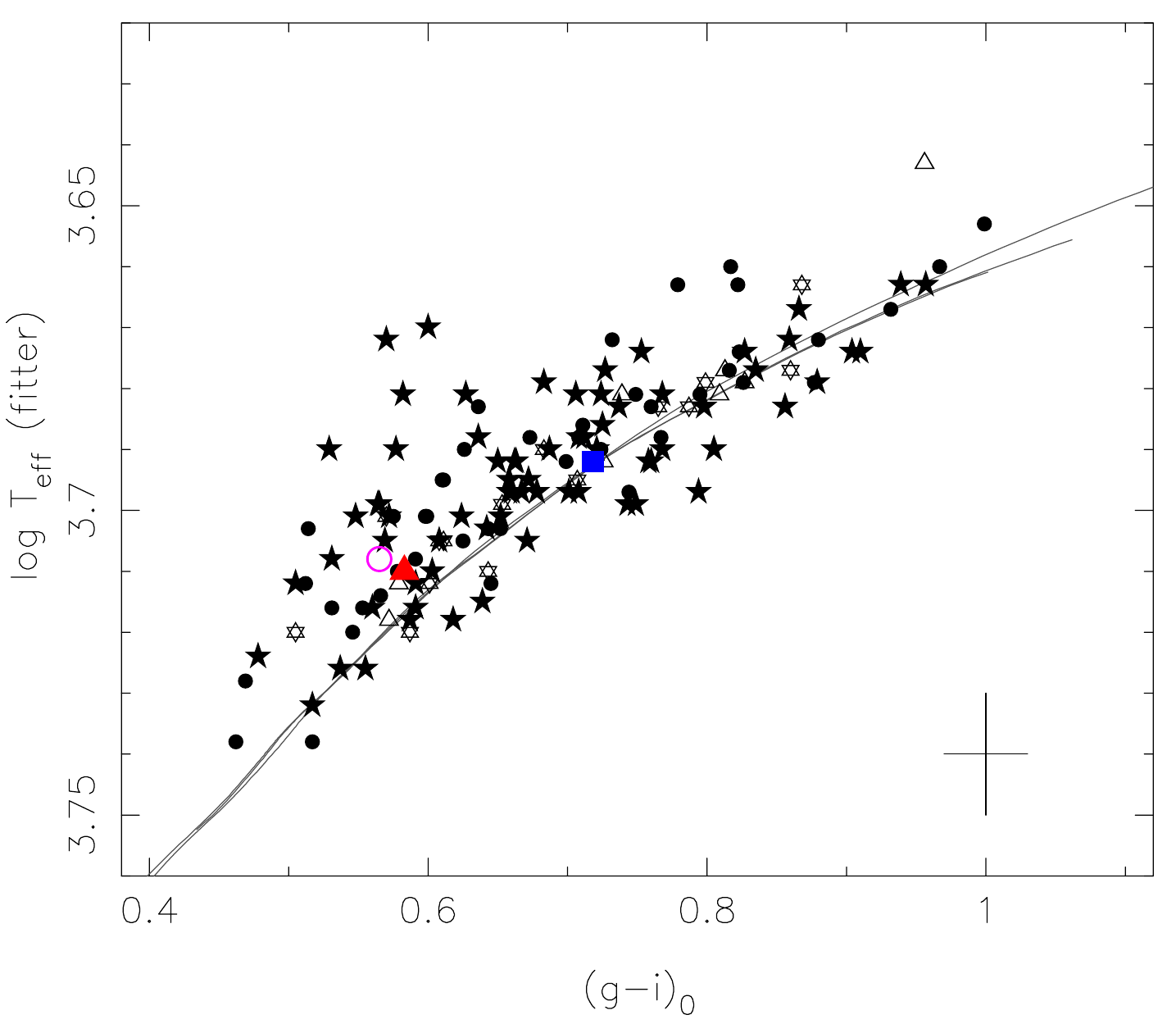}
\caption{The relation between the log $T_{\rm eff}$ and $(g-i)_0$ for the 142 stars with [Fe/H]$_{fitter}$ $<$ --3.0 dex.  Symbol code is as for Fig.\ \ref{logTeff-logg}.  Shown also are the relations for  metal-poor isochrones with [M/H] = --4.0, --3.5, --2.5 and 
[$\alpha$/Fe] = +0.4 for an age = 12.5 Gyr.  $\pm$1$\sigma$ error bars for both quantities are shown in the lower right of the panel.}
\label{gmi-logTeff}
\end{figure}

The relationship between the {\it fitter} log $T_{\rm eff}$ values and the SkyMapper DR1.1 $(g-i)_0$ colours is shown in Fig.\ \ref{gmi-logTeff}.  Shown also on the figure are the colour-temperature relations for the [M/H] = --4.0, --3.5 and --2.5 isochrones with
[$\alpha$/Fe] = +0.4 and age = 12.5 Gyr (see \S \ref{knownEMP}); the isochrones are quite consistent with the observed values.  A parabolic least-squares fit to the data has an {\it rms} of 0.01 in log $T_{\rm eff}$ which, for the median log $T_{\rm eff}$ of 3.692, corresponds to a temperature uncertainty of $\pm$100 K.  As for uncertainty in the {\it fitter} log~$g$ values, the {\it rms} about a linear least-squares fit to the points in Fig.\ \ref{logTeff-logg} suggests that the uncertainty is 0.3--0.35 dex.  The 1$\sigma$ errors in the
individual $(g-i)_{0}$ values are of order $\pm$0.03 mag (see \S \ref{phot_sect}).

\subsection{Metallicity Comparisons}

In order to judge the utility of the [Fe/H]$_{fitter}$ values for selecting stars to follow-up with high-dispersion spectrographs on 8m class telescopes, we have compared the [Fe/H]$_{fitter}$ values with those in the literature for a set of stars whose abundances have been determined from the analysis of high-dispersion spectra.  This set of 30 stars includes both known stars that were deliberately observed at the 2.3m telescope for this purpose, as well as those that represent the serendipitious rediscovery of known stars among the set of SkyMapper EMP candidates observed.

\begin{figure}
\centering
\includegraphics[angle=0.,width=0.48\textwidth]{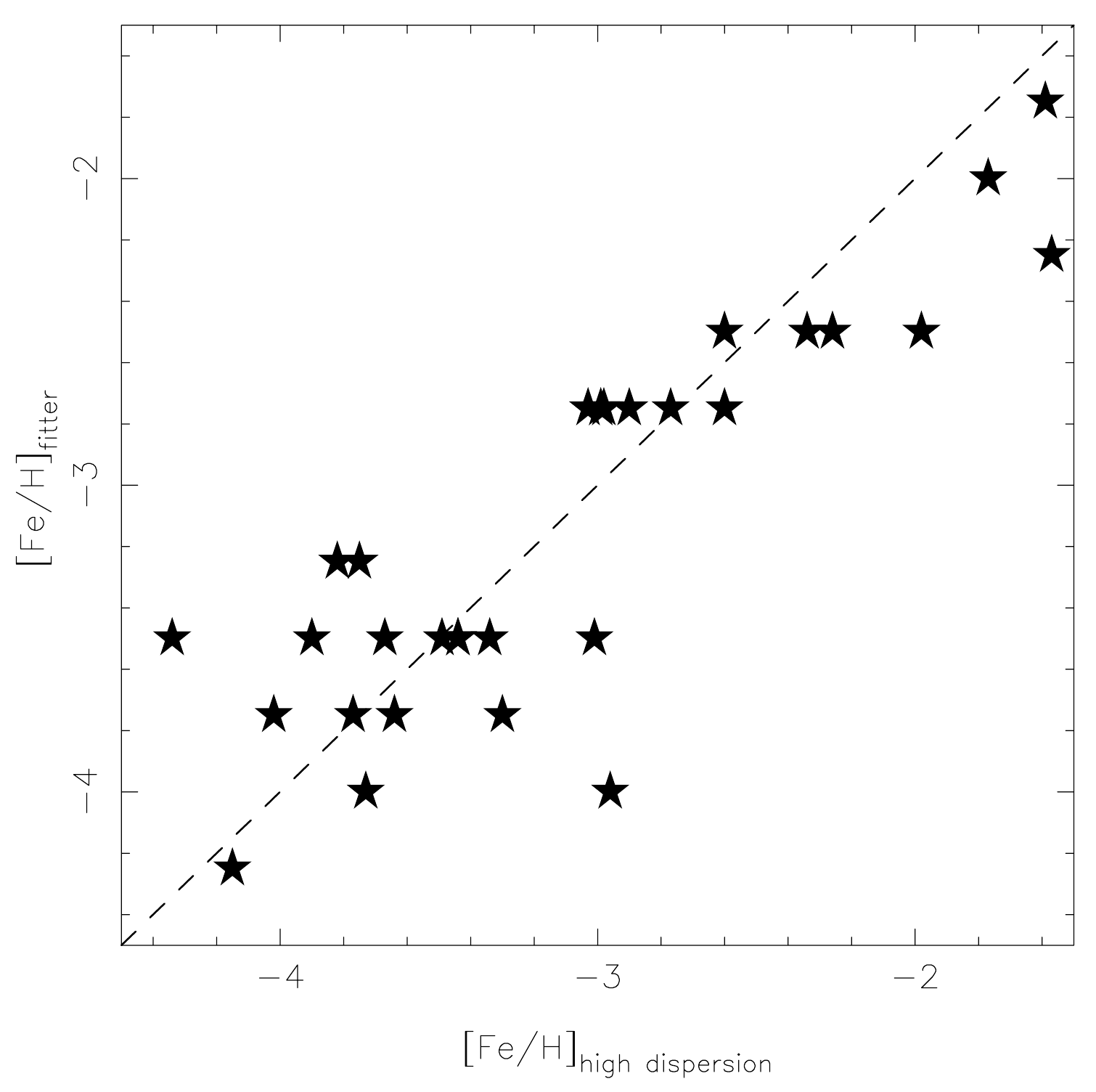}
\caption{The relation between [Fe/H]$_{high~dispersion}$, which are literature values derived from high-dispersion spectroscopy, and [Fe/H]$_{fitter}$, the values derived from the spectrophotometric fits to the 2.3m low resolution WiFeS spectroscopy.  The dashed-line is the 1:1 relation.  }
\label{feh_comp_fig}
\end{figure}

The results are shown in Fig.\ \ref{feh_comp_fig} and are given in Table \ref{known_stars_tab} and in the online supplementary material, together with the sources of the literature [Fe/H] values, which are those from 1D, LTE analyses.  We note that here we have not attempted to correct the literature [Fe/H]  values for the potentially different $T_{\rm eff}$ scales adopted in the cited papers.   Nevertheless, for these 30 stars the mean difference, in the sense literature [Fe/H]$_{high\ dispersion}$ minus [Fe/H]$_{fitter}$, is 0.04 $\pm$ 0.07 with a standard deviation of 0.38 dex.  Given that the [Fe/H]$_{fitter}$ values are quantized at the 0.25 dex level, the lack of any systematic difference is reassuring, while the spread suggests that a typical uncertainty in [Fe/H]$_{fitter}$ is of order 0.3 dex.  


Overall Fig.\ \ref{feh_comp_fig} suggests that  selecting stars with [Fe/H]$_{fitter}$ $<$ --3.0  for high-dispersion spectroscopic follow-up is likely to result in a high rate of return.  Table \ref{tab:data}, and the online supplementary material, list the SkyMapper designation, J2000 position, $g$, ($g-i$)$_{0}$ and metallicity index $m_{i}$ from the DR1.1 photometry, G-band (CH) equivalent width (see \S \ref{C_abund}), and the spectrophotometric fit parameters for the 142 stars with [Fe/H]$_{fitter}$ $<$ --3.0 dex.  High dispersion spectroscopic follow-up is presented and discussed in \citet{TN18} for the most metal-poor object 
SMSS J160540.18--144323.1, and in \citet{DY18} 
for a large fraction of the other stars in Table \ref{tab:data} based on observations obtained with the MIKE echelle spectrograph on the Magellan Clay 6.5m telescope.  For completeness we give in Table \ref{tab:data2}, and in the online supplementary material, the same information as for Table \ref{tab:data} for the 29 stars observed at the 2.3m which have \mbox{[Fe/H]$_{fitter}$ $\leq$ --3.25} that either fail the DR1.1 photometric selection criteria and/or lie outside the adopted selection window.


\begin{table*}
\caption{Comparison of [Fe/H]$_{fitter}$ values with those from high-dispersion spectroscopy literature values (complete Table available electronically).}
\label{known_stars_tab}
\begin{center}
\begin{tabular}{lcclc}
\hline
ID & [Fe/H]$_{high\ dispersion}$ & [Fe/H]$_{fitter}$ & Reference & Notes \\
\hline
SDSS J1024+1201 & --4.34 & --3.50 & Placco et al. (2015) & \\
CD --38$^{o}$ 245 & --4.15 & --4.25 & \citet{DY13} & \\
HE 2139--5432 & --4.02 & --3.75 & \citet{DY13} &  1 \\
HE 2302--2154 & --3.90 & --3.50 & Holleck et al.\ (2011) & \\
HE 0013--0256 & --3.82 & --3.25 & Holleck et al.\ (2011) & \\
CS22172--002 & --3.77 & --3.75 & \citet{DY13} & \\
HE 0926--0546 & --3.73 & --4.00 & Cohen et al.\ (2013) & \\
HE 2318--1621 & --3.67 & --3.50 & Placco et al.\ (2015) \\
SDSS J1322+0123 & --3.64 & --3.75 & Placco et al.\ (2015) \\
SMSS J033005.90--681352.5 & --3.44 & --3.50 & Jacobson et al.\ (2015)\\
SMSS J125804.65--335045.1 & -3.44 & --3.50 & Jacobson et al.\ (2015)\\
HE 0305--5442 & --3.30 & --3.75 & Cohen et al.\ (2013) & \\
2MASS J12570460--1054072 & --2.96 & --4.00 & Schlaufman \& Casey (2014) \\
\hline
\end{tabular}
\end{center}
Notes:
(1) Star has [C/Fe] $\approx$ +2.6, fit to the 2.3m spectrum modified to exclude strong CH features\\

\end{table*}

\begin{table*}
\caption{Details of the derived parameters for the observed stars with [Fe/H]$_{fitter}$ $<$ --3.0 (complete Table available electronically).}
\label{tab:data}
\begin{center}
\begin{tabular}{ccccccccccc}
\hline
ID & RA & Dec & $g$ & ($g-i$)$_{0}$ & $m_{i}$ & $W(G)$ & $T_{\rm eff}$ & log g & [Fe/H]$_{fitter}$ & Notes \\
    & (J2000) & (J2000) & & & & (\AA) &  & &  & \\
\hline
SMSS J160540.18--144323.1 & 16 05 40.18 & --14 43 23.1 & 15.999 & 0.718 & --0.106 & ~~7.09 & 4925 & 2.0~~~ & --4.75 & 1 \\
SMSS J054650.97--471407.8 & 05 46 50.97 & --47 14 07.8 & 14.066 & 0.583 & --0.168 & ~~1.95 & 5125 & 1.75~~ & --4.5~ & 2 \\
SMSS J205001.91--661329.7 & 20 50 01.91 & --66 13 29.7 & 13.298 & 0.565 & --0.122 & ~~2.84 & 5100 & 2.125 & --4.25 & 2 \\
SMSS J230702.23--693718.5 & 23 07 02.23 & --69 37 18.5 & 14.558 & 0.809 & --0.040 & ~~2.68 & 4800 & 1.125 & --4.1~ & 2 \\
SMSS J081112.13--054237.7 & 08 11 12.13 & --05 42 37.7 & 15.859 & 0.739 & --0.128 & ~~0.28 & 4800 & 1.75~~ & --4.0~ & 2 \\
SMSS J102731.70--411130.5 & 10 27 31.70 & --41 11 30.5 & 15.830 & 0.956 & --0.140 &  ~~2.06 & 4400 & 1.5~~~ & --4.0~ & 2 \\
SMSS J182845.79--354022.8 & 18 28 45.79 & --35 40 22.8 & 14.356 & 0.726 & --0.095 & --0.01 & 4925 & 1.5~~~ & --4.0~ & 2 \\
SMSS J185624.55--421733.7 & 18 56 24.55 & --42 17 33.7 & 14.235 & 0.827 & --0.056 & ~~1.33 & 4775 & 1.0~~~ & --4.0~ & 2 \\
SMSS J194222.58--481437.8 & 19 42 22.58 & --48 14 37.8 & 15.084 & 0.813 & --0.106 & ~~3.66 & 4750 & 1.5~~~ & --4.0~ &  \\
SMSS J211747.91--404512.2 & 21 17 47.91 & --40 45 12.2 & 14.970 & 0.579 & --0.088 & ~~1.49 & 5150 & 2.125 & --4.0~ & 2 \\
\hline
\end{tabular}
\end{center}
Notes:
(1) High dispersion spectroscopic follow-up in \citet{TN18}\\
(2) High dispersion spectroscopic follow-up in \citet{DY18}

\end{table*}

\begin{table*}
\caption{Details of the derived parameters for additional candidates with [Fe/H]$_{fitter}$ $<$ --3.0 (complete Table available electronically).}
\label{tab:data2}
\begin{center}
\begin{tabular}{ccccccccccc}
\hline
ID & RA & Dec & $g$ & ($g-i$)$_{0}$ & $m_{i}$ & $W(G)$ & $T_{\rm eff}$ & log g & [Fe/H]$_{fitter}$ & Notes \\
    & (J2000) & (J2000) & & & & (\AA) &  & &  & \\
\hline
SMSS J084457.17--264325.3 & 08 44 57.17 & --26 43 25.3 & 15.489 & 0.715 & --0.070 & ~~1.54 & 4875 & 1.5~~~ & --3.75 & 1 \\
SMSS J091716.23--273716.5  & 09 17 16.23 & --27 37 16.5 &  15.184 &  0.944 & --0.003 &  ~~4.06  & 4225  & 1.5~~~  & --3.75 & 1 \\
SMSS J163040.08--715639.0 &  16 30 40.08 & --71 56 39.0 &  14.943 &  0.492 & ~0.020 &  ~~1.54 &  5250 &  2.25~~ & --3.75 & 1 \\
SMSS J173002.48--532901.2 &  17 30 02.48 & --53 29 01.2 &  14.775 &  0.773 & --0.067 &  ~~0.74 &  4675  & 1.125 & --3.75 &  1\\ 
SMSS J092357.06--203851.2 &  09 23 57.06 & --20 38 51.2 &  15.418 &  0.527 & --0.004 &  ~~3.99  & 5075  & 2.125 & --3.5  & 1 \\
SMSS J101305.43--372044.4 &  10 13 05.43 & --37 20 44.4 &  16.026 &  0.239 &  ~0.153 & ~~1.29 &  5600 &  1.5~~~ &  --3.5  & 2 \\
SMSS J102049.27--481043.1 &  10 20 49.27 & --48 10 43.1 &  12.435 &  0.143 &  ~0.226 &  ~~1.55 &  5300 &  1.75~~ & --3.5 &  2 \\
SMSS J134416.03--235714.9  & 13 44 16.03 & --23 57 14.9  & 16.058  & 0.416 & --0.030 &  ~~0.90 &  5400  & 1.5~~~ &  --3.5 &  2 \\
SMSS J154634.19--081030.9 &  15 46 34.19 & --08 10 30.9  & 15.094 &  0.583 & --0.009 &  ~~1.20 &  5050 &  2.125 & --3.5 &  1\\ 
SMSS J181200.10--463148.7 &  18 12 00.10 & --46 31 48.7 &  13.837 & 0.850 & --0.112  & ~~1.92 &  4650 &  1.0~~~ &  --3.5 &  1 \\
\hline
\end{tabular}
\end{center}
Notes:
(1) Fails one of more DR1.1 photometric selection criteria~~~~~~~~~~~~~~~~~~~~~~~~~~~~~~~~~~~~~~~~~~~~~~~~~~~~\\
(2) Lies outside selection window (may also have failed the photometric selection criteria)

\end{table*}

\subsection{Selection Efficiency}

Of the 2618 individual EMP-candidates meeting the SkyMapper DR1.1 photometric database requirements that fall within the adopted selection window, i.e., have a photometric metallicity $\leq$ --2.0 (see \S \ref{ASW}), and which also have a least one 2.3m spectrophotometric observation, 1081 or 41.4\% have [Fe/H]$_{fitter}$ $\leq$ --2.75, while 485 (18.6\%) have [Fe/H]$_{fitter}$ $\leq$ --3.0 dex.  Overall 93\% of the stars observed have [Fe/H]$_{fitter}$ $\leq$ --2.0,  and only 6.7\% have [Fe/H]$_{fitter}$ $>$ --2.0, i.e., are true contaminants.   These figures indicate a high selection efficiency for metal-poor stars, supporting the original premise of adding the $v$-filter to the SkyMapper survey filter set.  Comparisons of efficiency to other previous and on-going large-area surveys \citep[e.g.][]{Beers92, Frebel06, Christlieb08, Li2015, SC14} are not straightforward as the base samples are often not well-defined.  Nevertheless, some comparisons are warranted.  For example, for the objective prism based HK-survey, \citet{Beers92}  report that intermediate-resolution spectroscopic follow-up reveals $\sim$7\% of stars selected as metal-poor candidates have [Fe/H] $\leq$ --3 dex.  Similarly, \citet{SC14} indicate, based on both low and high resolution spectroscopic follow-up, that approximately 4\% of their EMP candidates, which are selected using optical, near- and mid-infrared photometry, have [Fe/H] $\leq$ --3.0 dex.  

The most appropriate comparison as regards our selection efficiency is that with the results of the Pristine survey \citep{Starkenburg17}, which employs photometry from a narrow-band filter centered on the Ca {\sc ii} H and K-lines combined with SDSS broad-band photometry to define a photometric metallicity [Fe/H]$_{Pristine}$.  \citet{Starkenburg17} estimate their efficiency by comparing the photometric [Fe/H]$_{Pristine}$ estimates with SDSS/SEGUE metallicities [Fe/H]$_{SSPP}$ that are derived from low-resolution spectra \citep[e.g.][]{Lee2008}.  They find that 24\% of stars with [Fe/H]$_{Pristine}$ $\leq$ --3.0 also have [Fe/H]$_{SSPP}$ $\leq$ --3.0 and that 90\% of stars with [Fe/H]$_{Pristine}$ $\leq$ --2.5 have [Fe/H]$_{SSPP}$ $\leq$ --2.0 dex.  Such efficiencies are quite similar to those found here.

\citet{Youakim17} have also discussed the efficiency of the Pristine survey through directly obtaining low-resolution spectroscopy
of a sample of Pristine stars that were selected to meet specific criteria \citep[see][for details]{Youakim17}.  In a similar process to that followed here, temperatures, gravities and metal abundances were derived from the low-resolution spectra via comparisons with a 
grid of synthetic spectra
\citep{AP14,Youakim17}.  \citet{Youakim17} find that for the 130 stars with low-resolution spectra for which the Pristine photometry predicts [Fe/H] $\leq$ --2.5, 91 or 70\% have [Fe/H]$_{low-res}$ $\leq$ --2.5, and of these 10 or 8\% have [Fe/H]$_{low-res}$ $\leq$ --3.0 dex.  Similarly,
\citet{Youakim17} note that for the 46 stars predicted by Pristine photometry to have Fe/H] $\leq$ \mbox{--3.0}, 10 stars were found to have 
[Fe/H]$_{low-res}$ $\leq$ \mbox{--3.0} representing an efficiency of 22\%.  These efficiences are again similar to those for the SkyMapper sample.

Of course it must be kept in mind that the selection efficiency is directly related to the selection window -- in this study, because of our deliberate choice of requiring $(g-i)_{0}$ $\geq$ 0.4 for the spectroscopic follow-up of EMP-candidates, we have effectively excluded EMP-stars near the turnoff, resulting in a sample that is very much dominated by giants.  This is in contrast to surveys such as the ToPos survey \citep{Caffau13}, for example, which is specifically focussed on EMP-candidates in the vicinty of the main sequence turnoff.  

\subsection{Carbon abundance estimates} \label{C_abund}

As is now well established, the carbon-to-iron abundance ratio [C/Fe] in metal-poor stars exhibits increasing diversity as [Fe/H] decreases \citep[e.g.][]{DY13,Placco14}.  Two classes of objects are conventionally recognised -- one, the CEMP stars, which have [C/Fe] $\geq$ +0.7 
\citep{Aoki07} and for which there are a number of sub-classes \citep[see, for example,][for the definitions]{Placco14}, and, two, the stars which have [C/Fe] $<$ +0.7 dex that we will refer to as ``carbon-normal''.   The CEMP-stars are dominant at the lowest metallicities \citep[e.g.][]{Placco14,Yoon18}.  Consequently, we have investigated the [C/Fe] values for the stars with abundances [Fe/H]$_{fitter}$ $<$ --3.0 dex.  

\begin{figure}
\centering
\includegraphics[angle=0.,width=0.48\textwidth]{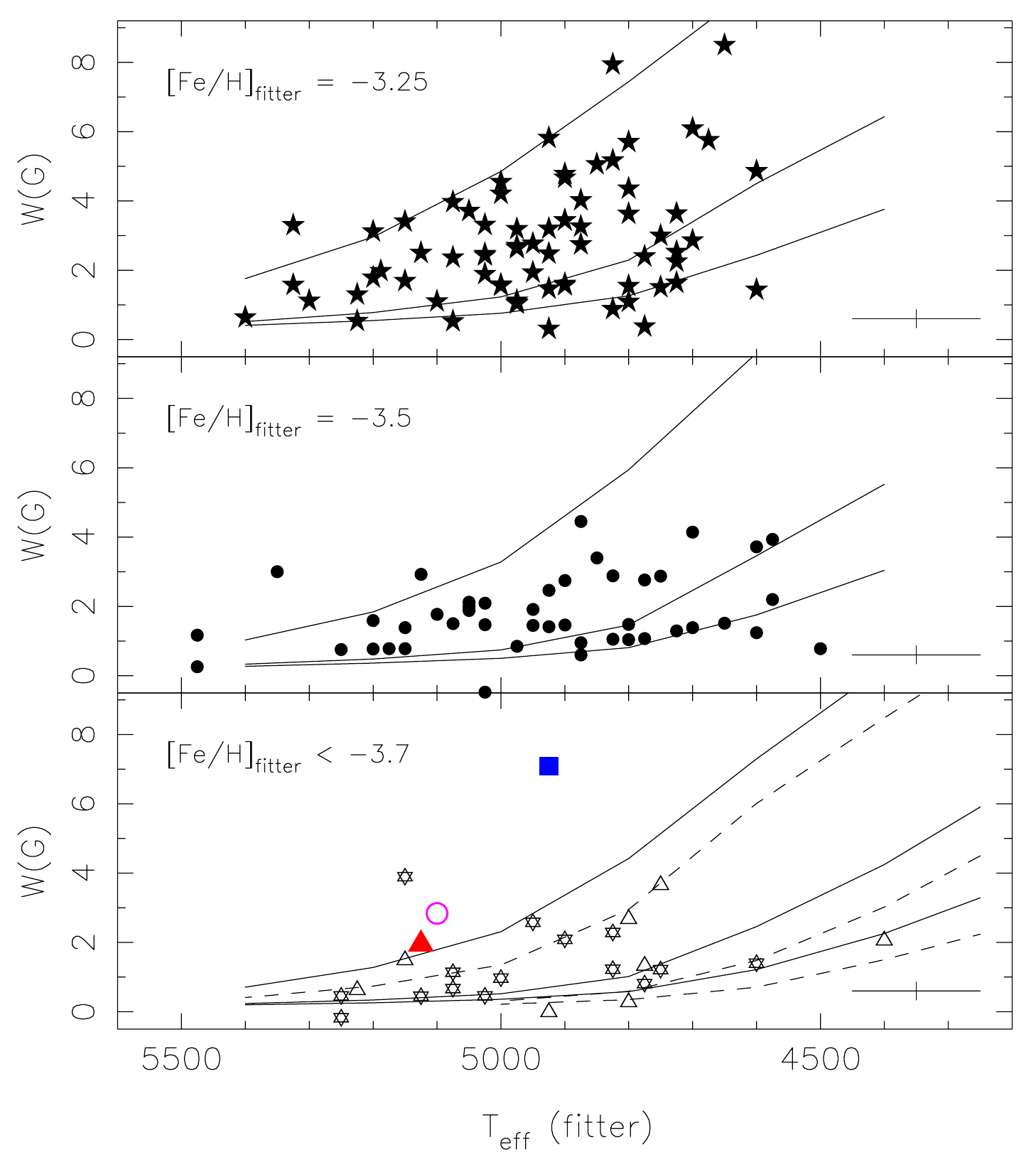}
\caption{Observed G-band (CH) feature strength $W(G)$, in \AA, as a function of the $T_{\rm eff}$ determined from the spectrophotometric fits.  Symbol code is as for Fig.\ \ref{logTeff-logg}, noting that the three stars with the lowest [Fe/H]$_{fitter}$ values are represented by the blue filled square, the red filled triangle and the magenta open circle.  All are likely C-rich objects.  1$\sigma$ error bars ($\pm$100 K, $\pm$0.3\AA)  are shown in the lower right of each panel.  Each panel also shows the relation between $W(G)$ and $T_{\rm eff}$ for the median log $g$ of each metallicity group and [C/Fe] = --0.5, 0.0 and +1.0, respectively, calculated from synthetic spectra.  The [Fe/H] values used are \mbox{--4.0} (dashed lines) and --3.75 (solid lines) in the lower panel, --3.5 for the middle panel, and --3.25 for the upper panel. }
\label{Gband_fig}
\end{figure}

To do this we measured an index, denoted by $W(G)$, on the velocity-corrected continuum-normalized spectra for the 142 stars in the sample with [Fe/H]$_{fitter}$ $<$ --3.0 dex.  Here $W(G)$ is the pseudo-equivalent width of the CH G-band feature obtained by numerically integrating the residual flux between the wavelengths  
$\lambda$4277--4318\AA\/ relative to the continuum defined by the regions $\lambda$4230--4265\AA\/ and $\lambda$4410--4440\AA.  The results are shown in the panels of Fig.\ \ref{Gband_fig} where we have separated the stars into three metallicity groups: [Fe/H]$_{fitter}$ $\leq$ --3.75 (lower panel), [Fe/H]$_{fitter}$ = \mbox{--3.5} (middle panel) and [Fe/H]$_{fitter}$ =  --3.25 (upper panel).  Based on stars with multiple 2.3m WiFeS spectra, the typical error in the individual $W(G)$ values, shown in the lower right section of each panel, is small: 1$\sigma$ $\approx$ 0.3\AA.  

What is apparent from the figure is that, not unexpectedly, there is a large range in $W(G)$ values in each metallicity group.  While some of the scatter is likely due to errors in the spectrophotometric temperatures and in the [Fe/H]$_{fitter}$ values, it is clear that there is a large intrinsic range in the $W(G)$ values at fixed 
$T_{\rm eff}$ and therefore, most probably, a substantial range in [C/Fe] amongst these stars.   Part of this variation is likely due to evolutionary mixing -- as a star ascends the red giant branch the convective envelope expands inwards reaching layers affected by CN-cycling.  The resulting mixing drives a reduction of the surface carbon abundance below the original natal abundance while increasing the surface nitrogen abundance.  The size of the evolutionary mixing effect on the surface abundance of carbon is a function of the star's $T_{\rm eff}$, log $g$ and [Fe/H] \citep[e.g.][]{Placco14} being largest for cool metal-poor stars near the RGB-tip.  

More specifically, using the tool available at {\it http://vplacco.pythonanywhere.com}, which implements the \citet{Placco14} corrections, the effects of evolutionary mixing can be estimated.  We find that the corrections to the observed abundances are only significant ($\Delta$[C/Fe] $\geq$ 0.3 dex) for [Fe/H] $\leq$ --3.25 if log $g$ $\leq$ 1.5.  The corrections range from $\sim$0.35 dex at log $g$ = 1.5 to $\sim$0.7 dex at log $g$ = 1.0.  However, since the majority of the stars with [Fe/H] $<$ --3.0 in our sample have log $g$ $\geq$ 1.5 (see Fig.\ \ref{logTeff-logg}) we have not attempted to apply the corrections. 

Nevertheless, while evolutionary mixing undoubtedly contributes to the intrinsic scatter in Fig.\ \ref{Gband_fig}, particularly at the cooler temperatures, it is likely that a substantial range in natal [C/Fe] values is present.  For example, the most metal-poor star in this sample, SMSS J160540.18--1414323.1, which is the blue filled square symbol in the bottom panel of Fig.\ \ref{Gband_fig} and which has one of the highest $W(G)$ values, is shown in \citet{TN18} to be strongly carbon-enhanced ([C/Fe]$_{1D,LTE}$ = 3.9 $\pm$ 0.2).  The other two stars with [Fe/H]$_{fitter}$ below --4 (the red filled triangle and the magenta open circle in Fig.\ \ref{Gband_fig}) also appear to be C-rich.

We illustrate the likely range in [C/Fe] present by making use of synthetic spectrum calculations: synthetic spectra were calculated, using the 1D, LTE approximation, for a range of $T_{\rm eff}$ values and for the median log $g$ for each set of [Fe/H]$_{fitter}$ values, i.e., [Fe/H] = --3.25, --3.5, --3.75 and \mbox{--4.0} \citep[see][]{TN18}.  The [C/Fe] values employed were --0.5, 0.0 and 1.0 dex.  The resulting synthetic spectra were then convolved to the observed resolution, continuum normalized in the same way as for the observed spectra and the resulting $W(G)$ indices measured.  The results are shown as the solid and dashed lines in Fig.\ \ref{Gband_fig}.  Evidently there is a substantial range in [C/Fe] among the stars in this sample, but aside from the most metal-poor star (blue filled square symbol) and the post-AGB star (the six-point star symbol at 5150, 3.9) there are no clear indications of any stars in the current sample having observed [C/Fe] significantly greater than +1.0 dex.  

\begin{figure}
\centering
\includegraphics[angle=0.,width=0.48\textwidth]{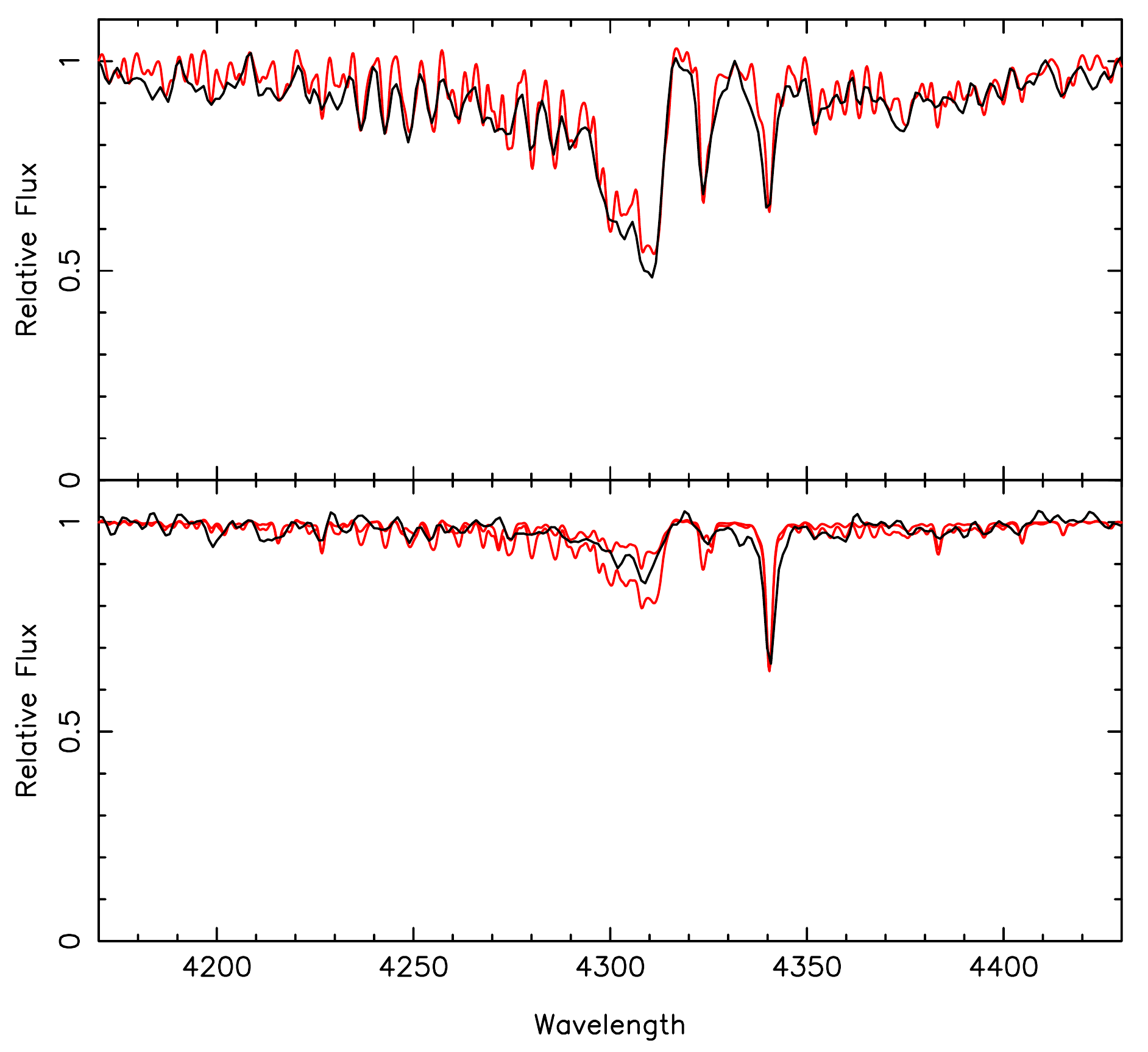}
\caption{A comparison of observed spectra with synthetic spectra in the vicinity of the G-band (CH).  {\it Upper panel:} observed spectrum (black line) for the star SMSS J191147.60--351911.7, for which the {\it fitter} determined values of $T_{\rm eff}$, log $g$ and [Fe/H] are 4825/1.625/--3.25 respectively, compared with a synthetic spectrum for 4800/1.5/--3.25 and [C/Fe] = +1.0 (red line).
{\it Lower panel:} Observed spectrum for the star SMSS J054650.97--471407.8 (black line), which has {\it fitter} values of  5125/1.75/--4.5, with synthetic spectra (red lines) for 4800/1.5/--4.5 and [C/Fe] = +1.0 and +1.5 dex.  }
\label{Gband_fits}
\end{figure}

Two examples are shown in Fig.\ \ref{Gband_fits}, noting that these are not detailed fits of the synthetic spectra to the observed spectra, rather they are simply a comparison of appropriate models with observed spectra.  The upper panel shows the star SMSS J191147.60--351911.7 ([Fe/H]$_{fitter}$ = --3.25, $W(G)$ = 7.93) for which the observed [C/Fe] is approximately +1, given that the CH-features for [C/Fe] = +1.5 are too strong compared with the observations.  The lower panel shows SMSS J054650.97--471407.8 ([Fe/H]$_{fitter}$ \mbox{= --4.5}, $W(G)$ = 1.95), which has observed [C/Fe] $\approx$ 1.2 dex, given the strengths of the CH features compared to the models at [C/Fe] = +1.0 and +1.5 dex.  Note that for this star the $T_{\rm eff}$ used for the synthetic spectra (4800 K) is somewhat cooler than the {\it fitter} value (5125 K) in order to match the strenght of the H$\gamma$ line in the observed spectrum.  In both cases the evolutionary mixing corrections to the observed carbon abundance are of order 0.35 dex (natal abundance higher).  Both are likely CEMP stars.  We have not calculated further spectral fits for [C/Fe] determinations as for a large fraction of the stars plotted in Fig.\ \ref{Gband_fig}, particularly for those in the lower two panels, detailed carbon abundances have been determined from Magellan/MIKE high dispersion spectra and are discussed in \citet{DY18}.


\subsection{The Observed Sample} \label{Obs_Samp}

\begin{figure}
\centering
\includegraphics[angle=0.,width=0.48\textwidth]{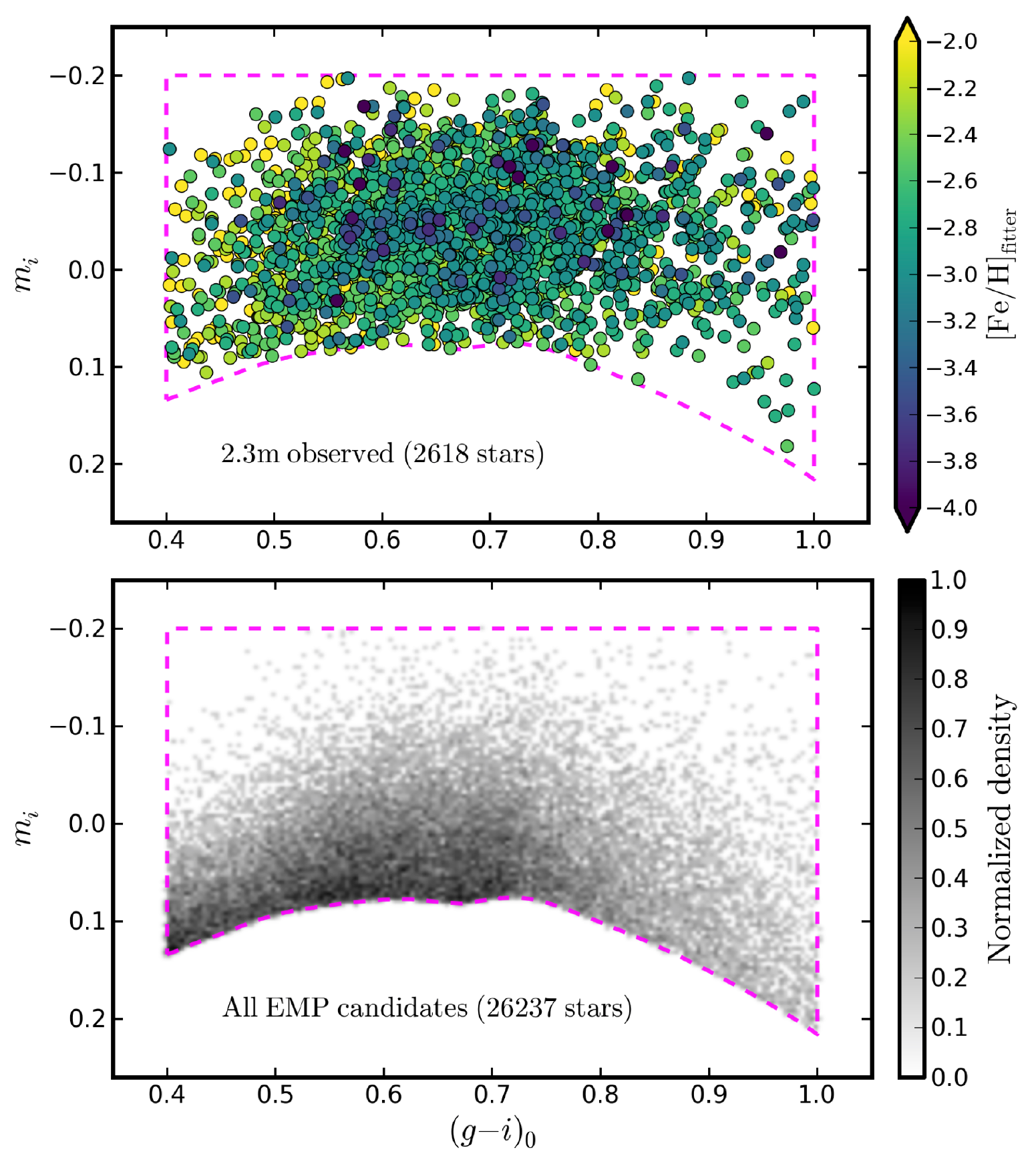}
\caption{Upper panel: the distribution within the adopted selection window, outlined by dashed lines, of the 2618 stars observed at the 2.3m.  The colour-coding for the derived fitter metallicities is shown at the right.  Lower panel: the corresponding distribution for the entire photometric sample of 26,237 stars.  }
\label{obs_stars}
\end{figure}

\begin{figure}
\centering
\includegraphics[angle=0.,width=0.48\textwidth]{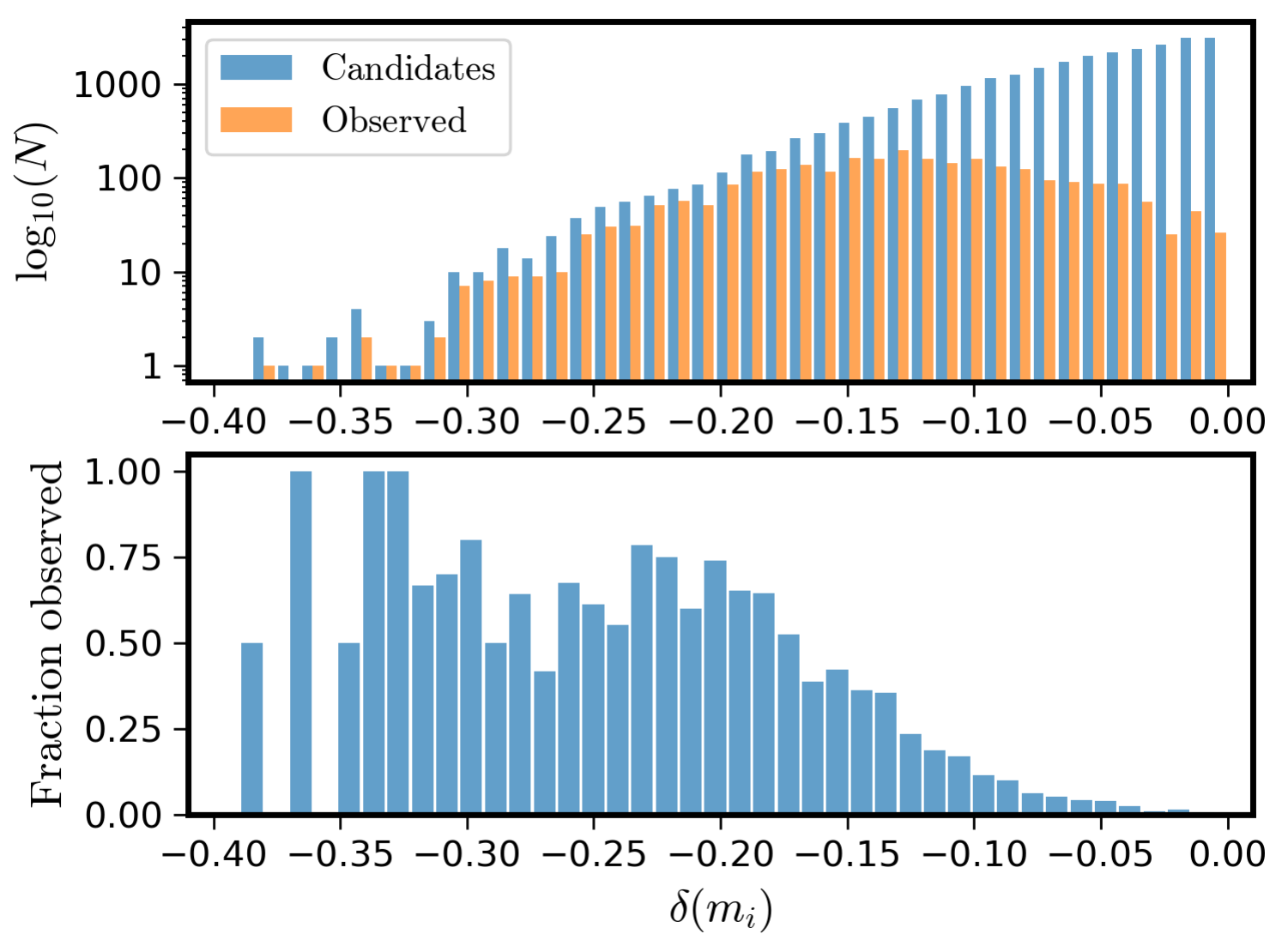}
\caption{Upper panel: histograms of the number of candidates observed as well as the number available as a function $\delta m_{i}$, the distance in magnitudes from the lower boundary of the selection window at the $(g-i)_{0}$ colour of the star.  Lower panel: the ratio of these two numbers, again as a function of $\delta m_{i}$.  }
\label{hist_obs_fig}
\end{figure}

The 2618 stars observed spectroscopically at low resolution represent almost exactly 10\% of the total sample of 26,237 stars in the adopted selection window.  The upper panel of Fig.\ \ref{obs_stars} shows the location within the selection window of these stars, while the lower panel shows the full photometric sample.  Not surprisingly, the distribution of the stars actually observed is clearly not a random sample of the underlying distribution --- the 2.3m observers have tended to select stars for observation that mostly have negative values of the metallicity index in the expectation that this will enhance the chances of finding extremely metal-poor stars.   This effect is illustrated in Fig.\ \ref{hist_obs_fig} where in the upper panel we show both the number of candidates observed and the number of candidates available as a function of $\delta m_{i}$, the difference in magnitudes at the $(g-i)_{0}$ of the star between the star's metallicity index and the value of $m_{i}$ for the lower boundary of the selection window.  The lower panel shows the ratio of these two quantities where it is evident that while we have at least 50\% completeness for $\delta m_{i}$ $\leq$ --0.17, the completeness drops rapidly as the number of candidates increases towards the lower boundary of the selection window.

We note that the observed sample is also not a random sample with respect to the distribution of $(g-i)_{0}$ colours --- the 2.3m observers have tended to avoid the bluest and reddest candidates, particularly for more positive $m_{i}$ values.  We assume that this latter bias does not introduce any systematic effects.  Similarly, as regards the magnitude distribution of the stars observed, again observers have shown a preference for the brighter candidates: 25.2\% of the $g$ $\leq$ 13.0 candidates have 2.3m spectra (219 stars observed), while for 13 $<$ $g$ $\leq$ 15 the observed fraction is 14.9\% (1409 stars) and for 15 $<$ $g$ $\leq$ 16.1 it is 6.2\% (990 stars).  

\begin{figure}
\centering
\includegraphics[angle=0.,width=0.48\textwidth]{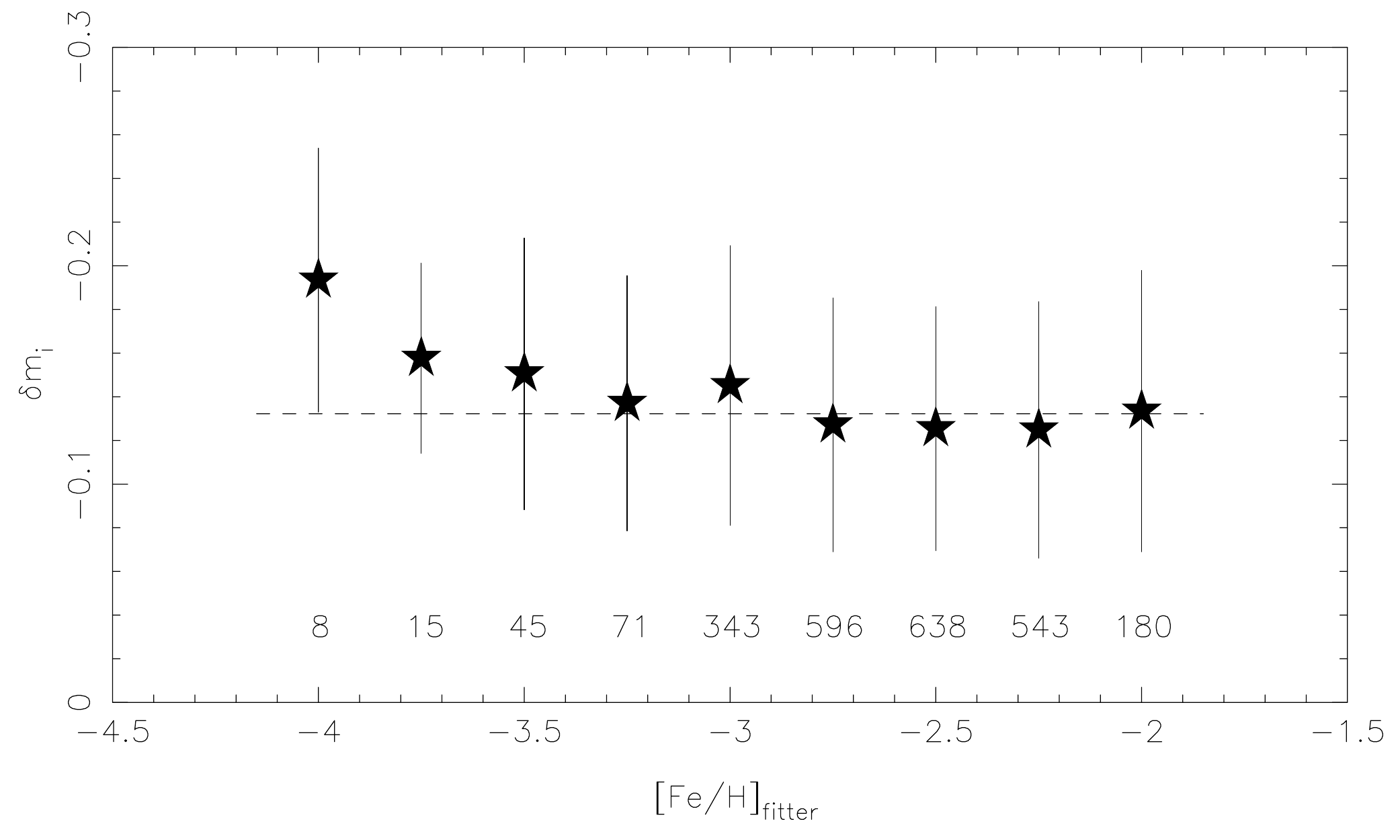}
\caption{The mean deviation in metallicity index $m_{i}$ from the base of the photometric selection window, $\delta m_{i}$, is shown 
as a function of [Fe/H]$_{fitter}$ for the observed stars with --4.0 $\leq$ [Fe/H]$_{fitter}$ $\leq$ --2.0 dex.  Shown also are the standard deviations in the $\delta m_{i}$ values.  The sample sizes for each metallicity bin are given beneath each point.  The dashed line is the mean value for the points with [Fe/H]$_{fitter}$ $\geq$ --3.25 dex.  
}
\label{delta_mi}
\end{figure}

The question then is: does this ``observer bias'' for more negative $m_{i}$ values result in the preferential discovery of more metal-poor stars than might be expected from the observation of a random (unbiased) selection of candidates from the input list?  The answer, perhaps unexpectedly, is basically ``no'' in the sense that there is only a minor increase in the relative yield of the lowest metallicity stars compared to more metal-rich candidates. 
We illustrate this by considering the relation between the $\delta m_{i}$ values and [Fe/H]$_{fitter}$; nominally the expectation is that $\delta m_{i}$ will be more negative for more metal-poor stars.   Fig.\ \ref{delta_mi} then shows the mean values of $\delta m_{i}$ for the sets of stars with  --4.0 $\leq$ [Fe/H]$_{fitter}$ $\leq$ --2.0 dex against 
[Fe/H]$_{fitter}$ values\footnote{Recall that the [Fe/H]$_{fitter}$ values are quantized at the 0.25 dex level.}.  Shown also on the plot are the standard deviations in the $\delta m_{i}$ values for each metallicity group.  

Fig.\ \ref{delta_mi} reveals that the stars with [Fe/H]$_{fitter}$ $\leq$ --3.5 do have slightly more negative mean $\delta m_{i}$ values than the stars with [Fe/H]$_{fitter}$ $\geq$ --3.25, but the difference is small: given the standard deviations, a specific
$\delta m_{i}$ value can correspond to basically any metallicity below --2.0 dex.  While in some sense this is a disappointing outcome (though it should be kept in mind that contamination by more metal-rich stars is very minor) it does mean that there is little metallicity bias in the observed sample, and any bias present is in the sense there are slightly more stars with [Fe/H]$_{fitter}$ $\leq$ --3.5 in the final sample than might be expected for a random sampling of the full candidate list.  The virtually constant standard deviation in the $\delta m_{i}$ values at $\sim$0.06 mag again demonstrates that we are currently limited by the errors in the $v$-band photometry.  

\begin{figure*}
\centering
\includegraphics[angle=0.,width=0.95\textwidth]{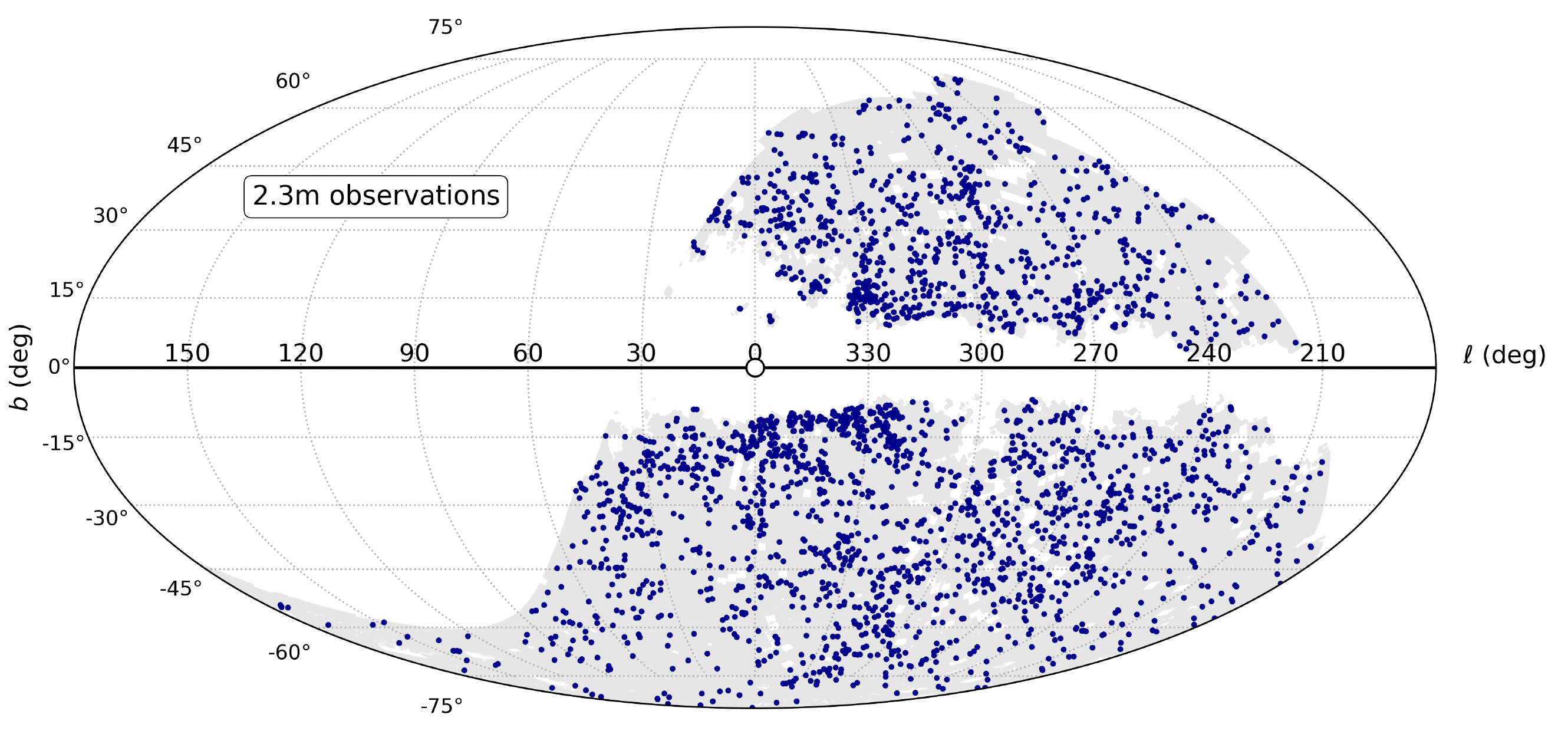}
\caption{As for Fig.\ \ref{allsky_fig} except that only the locations of the 2618 stars within the photometric selection window and which have been observed at the 2.3m are shown.   }
\label{cand_onsky}
\end{figure*}

For completeness we point out that the observed sample is also not consistent with a randomly selected sample as regards the distribution on the sky.  Observations are conducted year round, but as Fig.\ \ref{allsky_fig} shows, the full sample is strongly concentrated in the part of the sky towards the Galactic Centre.  Consequently, we have observed a smaller fraction of the total number of candidates in that direction than is the case, for example, at higher Galactic latitudes.  The on-sky distribution of all the candidates within the photometric selection window and which have been observed at the 2.3m is shown in Fig.\ \ref{cand_onsky}.  There does not appear to be any obvious dependence on metallicity in this distribution.

\section{Metallicity Distribution Function} \label{MDF}

\begin{figure}
\centering
\includegraphics[angle=0.,width=0.48\textwidth]{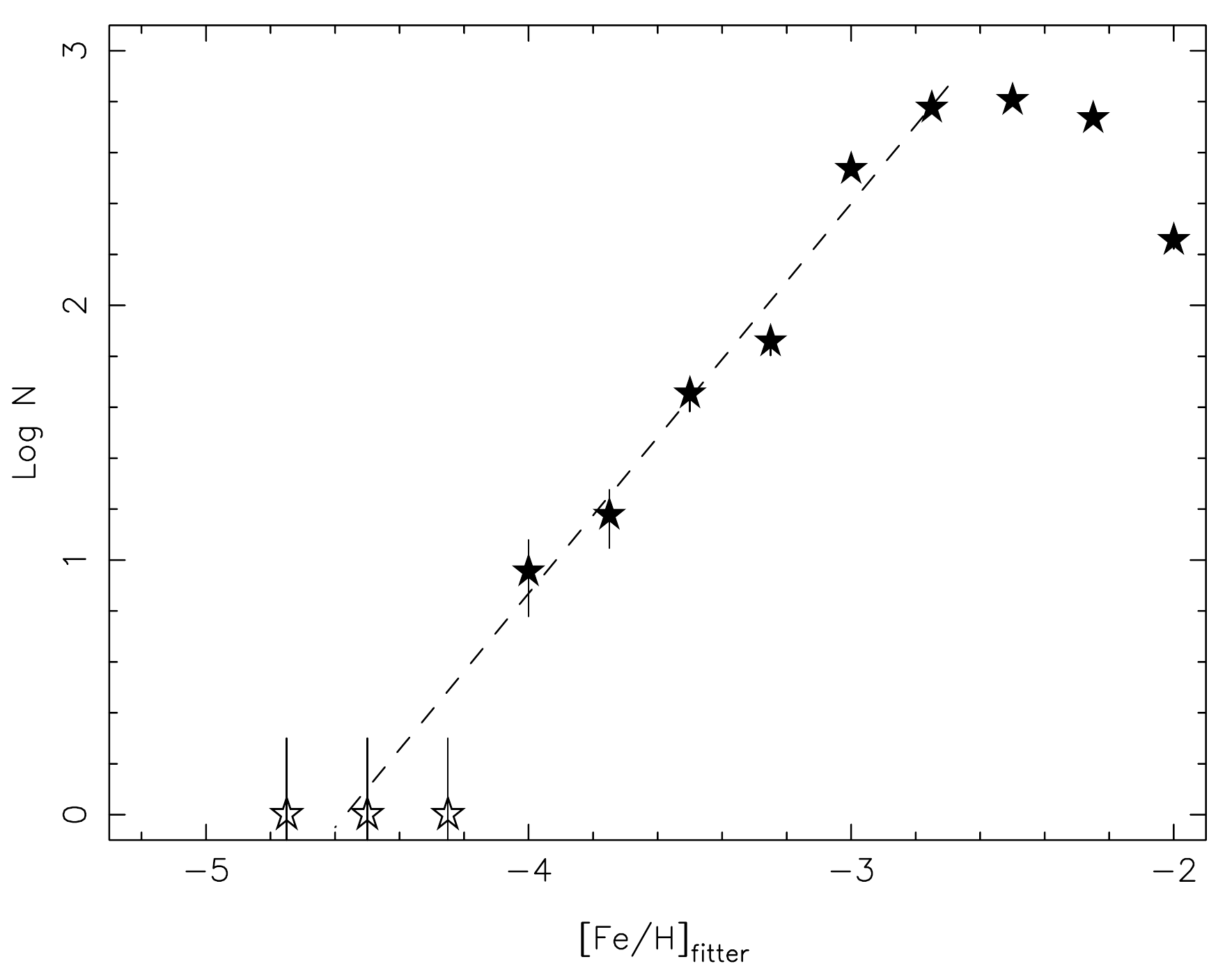}
\caption{The Metallicity Distribution Function (MDF) for the sample of stars within the photometric selection window with metallicities, [Fe/H]$_{fitter}$, determined from low resolution spectroscopy.  Error bars based on Poisson statistics are shown.  The dashed line is a least squares fit to the points between [Fe/H]$_{fitter}$ = --2.75 and --4.0 dex.  The slope is 1.5 $\pm$ 0.1 dex/dex.  The three lowest metallicity stars are plotted as open symbols to indicate that all three are likely C-rich objects.}
\label{MDF_fig}
\end{figure}

In Fig.\ \ref{MDF_fig} we plot the Metallicity Distribution Function (MDF) for the stars observed within the photometric selection window 
in the form of log~N versus 
[Fe/H]$_{fitter}$.  We emphasize that because of our colour-based primary candidate selection (0.4 $\leq$ $(g-i)_{0}$ $\leq$ 1.0; see \S \ref{ASW}) the MDF is dominated by giants.  What is immediately obvious from this figure is an apparent abrupt drop in the number  of stars below [Fe/H]$_{fitter}$ = --4.0 dex.  Above this metallicity up to the turnover at [Fe/H]$_{fitter}$ $>$ --2.75, the points are consistent with a power-law distribution, with slope $\Delta$(Log~N)/$\Delta$[Fe/H] = 1.5 $\pm$ 0.1 dex per dex.  This power slope does not change if the point at [Fe/H]$_{fitter}$ = --2.75 is excluded from the fit.

Extrapolating the power-law predicts a total of five stars should have been observed with --5.0 $\leq$ [Fe/H]$_{fitter}$ $\leq$ --4.25 whereas there are three such stars in the current sample, all of which are apparently C-rich (see Fig.\ \ref{Gband_fig}).  Therefore the statistical weight of the apparent drop in star numbers is low if the three C-rich lowest metallicity stars are included.  Indeed,  assuming the power-law slope remains valid, and no more stars with [Fe/H]$_{fitter}$ $\leq$ --4.25 are discovered, then a sample that is between 3 and 4$\times$ larger, i.e., more than a thousand stars at [Fe/H]$_{fitter}$ = \mbox{--3.0}, would be needed to establish the reality of the drop in the MDF at [Fe/H]$_{fitter}$ $<$ --4.0 with a significance exceeding the 3$\sigma$ level.  
However, as shown in Fig.\ \ref{Gband_fig}, the current sample is dominated by carbon-normal stars and there are no such stars in the sample with metallicities below --4 dex.  The difference between five predicted stars and zero observed stars is then a 2.2$\sigma$ result, which is somewhat more significant.  A sample size only a factor of two larger would then be sufficient to establish the drop with a significance exceeding the 3$\sigma$ level, if no carbon-normal stars with [Fe/H]$_{fitter}$ $\leq$ --4.25 occur in the 
enlarged sample.  For completeness we note that a similar drop in the MDF is seen 
at [Fe/H] $\approx$ --4.2 in the corrected (log~$\phi$) MDF discussed in \cite{DY13}, {\it which is a completely independent sample}.  The corrected (log~$\phi$) MDF in \citet{DY13} also has a power-law slope between [Fe/H] $\approx$ --4.0 and --3.0 of $\sim$1.7 dex per dex, similar to the value found here.    

Before speculating as to any physical interpretation of the apparent drop in the MDF at [Fe/H]$_{fitter}$ $<$ --4.0, however, it is necessary to investigate possible additional biases present in the observed sample.  Fig. \ref{delta_mi} indicates that there may be a minor bias towards that inclusion of lower metallicity stars in the sample, but accounting for it would only reduce the number of more metal-poor stars relative to the number at [Fe/H]$_{fitter}$ = --3.0 dex.  It is also necessary to keep in mind that the observed sample is likely a mixture of both inner and outer halo populations \citep{DC07,DC10,An13,Yoon18}, which may or may not have similar MDFs for [Fe/H] $\leq$ --3 and below: our sample likely includes stars at Galactocentric distances of up to $\sim$50 kpc.  We intend to perform a full kinematic and spatial analysis of the current sample in a later study.

The largest possible bias is the extent to which carbon-enhanced metal-poor (CEMP) stars, namely those with [C/Fe] $\geq$ +0.7 dex are, or are not, under-represented in the present sample.  In particular, CEMP stars, particularly CEMP-no stars (i.e., those lacking enhancements in $n$-capture elements) are increasingly common at abundances below [Fe/H] $\approx$ --3.0 and may in fact be dominant for [Fe/H] $<$ --4.0 dex \citep[e.g.][]{Placco14,Yoon18}.  Fig.\ \ref{Gband_fig} suggests that stars with large carbon enhancements are not common in our observed sample (except perhaps at the lowest metallicities) and so it may be that the drop in the observed MDF below --4.0 in Fig.\ \ref{MDF} is a result of missing a number of CEMP stars at and below this metallicity.  However, this is unlikely to be the entire explanation as the three stars in the sample with [Fe/H]$_{fitter}$ $<$ --4.0 and are all apparently C-rich (see Fig.\ \ref{Gband_fig}), and the so-called `commissioning survey', the pre-cursor of the current work, did discover SMSS J031300.36--670839.3, which has [Fe/H]$_{3D,NLTE}$ $<$ --6.5 \citep{TN17} and [C/H]$_{3D,LTE}$ = --2.55 \citep{MSB15}, i.e., [C/Fe] $>$ +4 dex.  For the most metal-poor star in the current sample, SMSS J160540.18--144323.1, which has 
[Fe/H$]_{fitter}$ $<$ --4.75\footnote{This star has E$(B-V)_{SFD}$ = 0.24 and as a result, the \mbox{Ca {\sc ii}} K line in the low resolution 2.3m spectrum, on which the {\it fitter} metallicity estimate is largely based, is likely to have a substantial interstellar component; the {\it fitter} value is therefore an upper limit on the actual abundance.}, \citet{TN18} in fact find [Fe/H]$_{1D,LTE}$ = --6.2 $\pm$0.2 and [C/Fe] $_{1D,LTE}$ = +3.9 $\pm$0.2 from the analysis of a high dispersion spectrum. 

We now discuss the role [C/Fe] plays in modifying, or not, the chance that carbon-enhanced stars are included in the observed sample of stars. 

\subsection{The role of [C/Fe]} \label{Cabund_sect}

\begin{figure}
\centering
\includegraphics[angle=0.,width=0.48\textwidth]{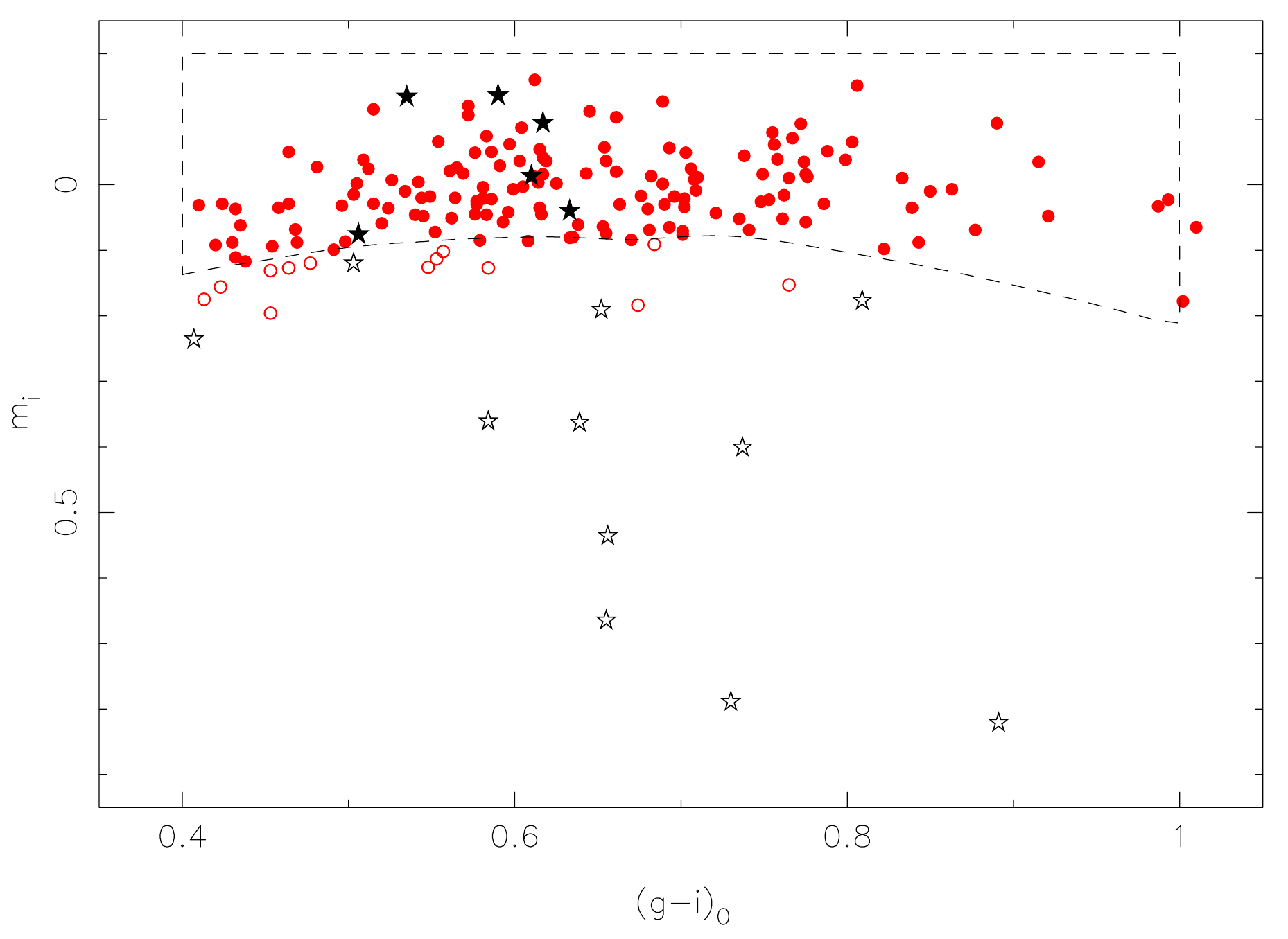}
\caption{The location in the SkyMapper metallicity-sensitive diagram of the 172 stars with [Fe/H] $\leq$ --2.5 dex based on high-resolution spectra and $(g-i)_{0}$ colours within the selection boundary intervals.  Open symbols are used for stars with $m_{i}$ values outside the selection box (shown by the dashed lines), filled symbols for stars within the selection box.  Black five-point star symbols are used for carbon-enhanced stars while red circles are used for carbon-normal stars.}
\label{known_fig}
\end{figure}

\begin{figure}
\centering
\includegraphics[angle=0.,width=0.48\textwidth]{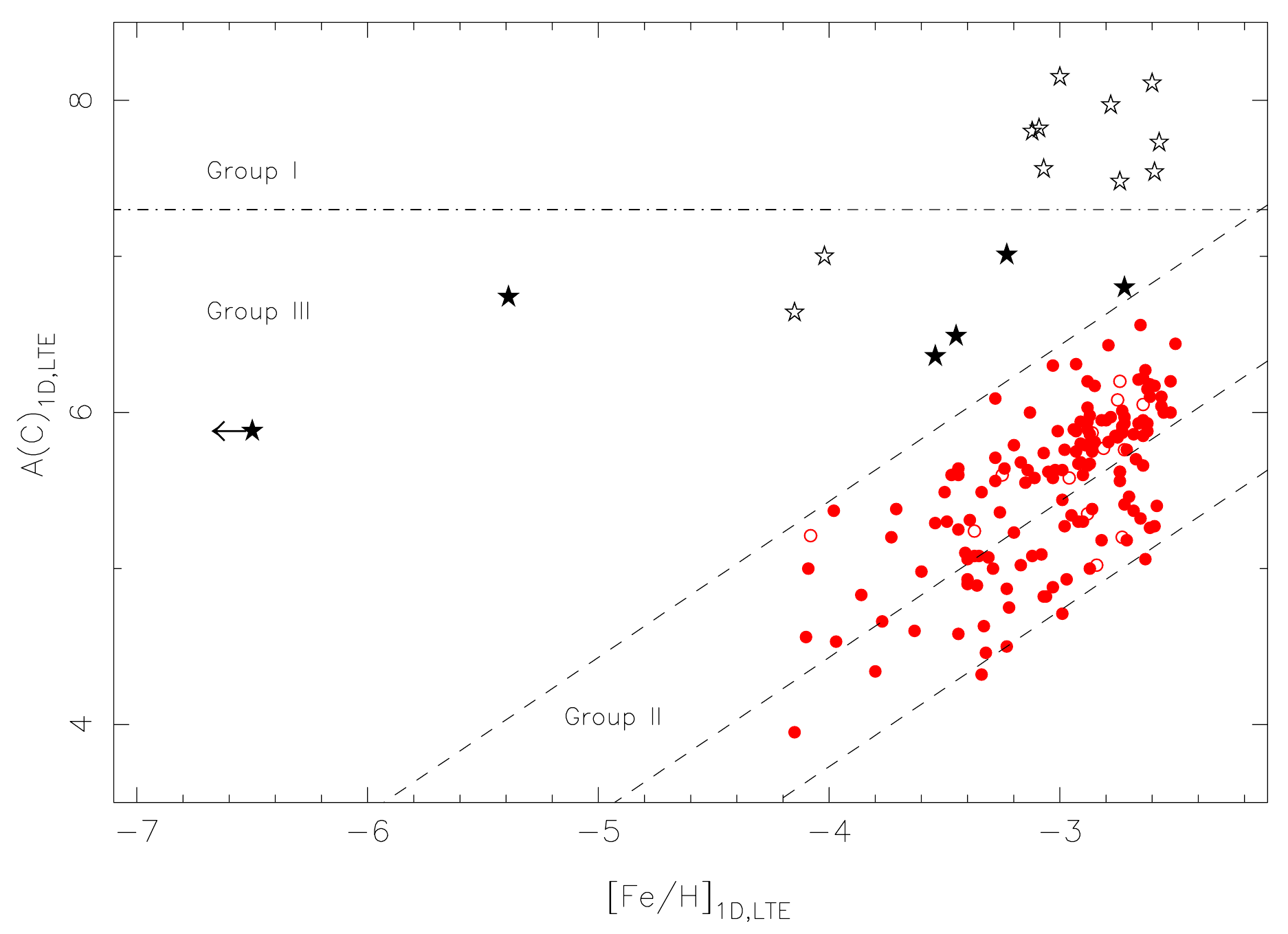}
\caption{Absolute carbon abundances A(C) plotted against [Fe/H] for the 172 stars shown in Fig.\ \ref{known_fig}, assuming a solar carbon abundance of 8.43.  Both abundances are based on 1D, LTE analyses.  As for Fig.\ \ref{known_fig}, black 5-pt star symbols are for carbon-enhanced stars (open if outside the selection window, filled if within) and red circles are for carbon-normal stars (open if outside the selection window, filled if within).  The dot-dash line at A(C) = 7.3 separates the \citet{Yoon16} `Group~I' stars from those in Groups II and III\@.  The diagonal dashed lines are for [C/Fe] = +1.0, 0.0 and --0.7, respectively.}
\label{ac_fig}
\end{figure}

In Fig.\ \ref{known_fig} we show the location in the SkyMapper metallicity-sensitive diagram of a subset of the 294 stars in our sample of known-EMP stars (see Fig.\ \ref{cmd_knownEMP}).  The subset consists of 172 stars which have 
both [Fe/H] and [C/Fe] derived from high
dispersion spectroscopic analyses that have [Fe/H] $\leq$ --2.5, that also have 0.4 $\leq$ $(g-i)_{0}$ $\leq$ 1.0 in the DR1.1 photometry database, and that meet our photometric seletion criteria.  In all cases the [Fe/H] and [C/Fe] values plotted are those from 1D, LTE analyses.  
In Fig.\ \ref{ac_fig} we also show the absolute carbon abundance A(C) against [Fe/H] for this same sample of 172 stars.
We note that there are no stars in common between this sample and the EMP-candidates with low resolution spectroscopy from the 2.3m telescope.

Given the morphology of Fig.\ \ref{ac_fig} where the [C/Fe] = +1 line forms a distinct upper bound to the set of stars for which [Fe/H] and A(C) vary together, we have elected to separate the stars into two classes, those with [C/Fe] $\leq$ +1, which we will refer to as the ``carbon-normal'' set (155 stars), and those with [C/Fe] $>$ +1, which we define as the CEMP stars category (17 stars)\footnote{We note specifically that the 
[C/Fe] values employed here are the observed values; we have not made any attempt to correct the carbon abundances for the effects of evolutionary mixing as discussed in \citet{Placco14}.}.  Our classification is different from the more conventional categorization of CEMP stars as those with [C/Fe] $\geq$ +0.7 \citep[e.g.][]{Aoki07}.  However, since the 7 stars in Fig.\ \ref{ac_fig} with [C/Fe] between +0.7 and 1.0 also follow the general correlation between A(C) and [Fe/H] exhibited by the stars with [C/Fe] $\leq$ 0.7, none of the subsequent discussion is affected by our choice of the [C/Fe] value to distinguish the two classes.

The expectation for the carbon-normal set is that they all should lie within the selection window, but in fact $\sim$8\% lie outside the window with $m_{i}$ values generally just below the lower boundary.  The metallicities of the stars outside the selection box do not show any obvious difference in distribution to that for the stars within the selection window: for example, the medians are [Fe/H] = --2.84 (n=13) and [Fe/H] = --2.93 (n=142), respectively.  
Examples of known carbon-normal stars whose photometry in the SkyMapper DR1.1 database places them outside our selection window include SMSS~J091210.40--064427.9 ([Fe/H] = \mbox{--2.64}) and SMSS~J133532.32--210632.9 ([Fe/H] = --2.73), both of which were identified as EMP-candidates in the SkyMapper commissioning-era photometry and followed-up at high dispersion by \citet{HJ15}.  The most metal-poor known-abundance carbon-normal star outside the current selection window is the star HE~0057--5959, with [Fe/H] = --4.09 \citep{DY13}, which was (re)discovered in the commissioning-era survey \citep{HJ15}.  We conclude that errors in the DR1.1 photometry, particularly, the $v$-band, mean that the use of the current selection window results in an underestimate of less than 10\% of the expected total number of carbon-normal stars with [Fe/H] $\leq$ --2.5 dex.  There does not, however, appear to be any metallicity bias induced by this effect.

The situation is more complex for the CEMP category.  For this grouping, six stars are within the window but eleven lie outside the window, with seven by a large margin (see Fig.\ \ref{known_fig}).  We address this by considering the absolute carbon abundance A(C) vs [Fe/H] diagram introduced originally by \citet{Sp13} and discussed extensively in \citet{Yoon16}.  The location of the 172 known-abundance stars in this diagram is shown in Fig.\ \ref{ac_fig}: symbols are the same as for Fig.\ \ref{known_fig}.  Unlike \citet{Yoon16} we have not restricted the stars plotted in the diagram to those with [C/Fe] $\geq$ +0.7, and this reveals that the `Group II' stars, as classified by \citet{Yoon16}, are simply the carbon-richer section of a sequence where A(C) and [Fe/H] vary together: lower [Fe/H] goes with lower A(C) with the scatter at fixed [Fe/H] likely a combination of evolutionary mixing effects (which can decrease A(C) from the original value) and intrinsic variation in the carbon abundances.  These are the ``carbon-normal'' stars referred to above.  For these stars, given the variety of sources from which they are drawn (see \S \ref{knownEMP}), there is no obvious evidence of incompleteness as a function of A(C) or [Fe/H], and this class clearly lacks stars more metal-poor than [Fe/H] $\approx$ --4.2, consistent with the MDF shown in Fig.\ \ref{MDF_fig}.

The carbon-rich stars in Fig.\ \ref{ac_fig} reveal a similar distribution to that seen in \citet{Yoon16}.  Nine have large A(C) values that exceed A(C) = 7.3 and these are classified as `Group I' stars in the terminology of \citet{Yoon16}.  \citet{Yoon16} note that the Group~I stars are predominantly CEMP stars that show enhancements of neutron-capture elements such as the CEMP-$s$ stars.  Indeed 8 of the 9 stars with A(C) $>$ 7.3 in Fig.\ \ref{ac_fig} are classified as \mbox{CEMP-$s$}, while one
\citep[CS22957-027, a known binary,][]{Hansen16} is a CEMP-no star.  None of these stars is within 0.1 mag in $m_{i}$ of the lower boundary of the selection window (see Fig.\ \ref{known_fig}) and most are 0.3 mag or more in $m_{i}$ away.  The obvious conclusion is that the approach adopted here to select EMP stars is very strongly biased against the selection of Group~I, i.e. predominantly CEMP-$s$, stars.

In the classification scheme of \citet{Yoon16}, stars that show little dependence of A(C) on [Fe/H], and which have A(C) $<$ 
$\sim$7, are known as Group~III stars.  There are eight stars in our known-EMP sample that are classifiable as Group~III stars  in 
Fig.\ \ref{ac_fig}.  Six of these stars are in the selection window including the two most Fe-poor stars: SMSS~031300.36--670839.3 with [Fe/H] $<$ --6.5 \citep{TN17} and HE~0107--5240 at [Fe/H] = --5.4 \citep{NC04}.  Both stars have [C/Fe] $>$ +3.5 dex.  The other four Group~III stars in the selection box are HE~1249--3121 ([Fe/H], [C/Fe]) = (--3.23, 1.81), HE~1351--1049 (--3.45, 1.51), HE~1506--0113 (--3.54, 1.47), and CS22877-001 \mbox{(--2.72, 1.09)}, respectively.  The two stars outside the selection window are HE~2139--5432 \mbox{(--4.02, 2.59)} and HE~1310--0536 (--4.15, 2.36), respectively.  HE~2139--5432 and HE~1249--3121 make an interesting pair:  the $(g-i)_{0}$ values are essentially identical at 0.503 and 0.506, and with $m_{i}$ = 0.076 HE~1249--3121 lies just inside the selection window, while at $m_{i}$ = 0.120 HE~2139--5432 lies just outside the selection window.  The stars also have essentially identical carbon abundance: [C/H] $\approx$ --1.4 dex.  Whether the difference $m_{i}$ for these two stars is real, or simply a consequence of errors in the $v$-band photometry cannot be definitely established.  However, we note that while the known-EMP stars used were specifically selected from DR1.1 to have low tabulated photometric errors, it is likely that systematic variations in the zero point of the $v$-band DR1.1 photometry across the sky are present.  Indeed, if the sample of 172 known abundance stars, less the 7 objects with $m_i$ $>$ 0.3 (see Fig.\ \ref{known_fig}), are grouped into 0.25 dex metallicity bins and the mean and sigma for the deviations from the base of the photometric selection box calculated (as in Fig.\ \ref{delta_mi}), then we again find that the standard deviation in each metallicity bin ($\sigma ( \delta m_{i})$ $\approx$ 0.06 mag) is virtually the same as for the stars observed at the 2.3m telescope.  Once more this indicates that there is an overall uncertainty in the DR1.1 $v$ magnitudes at the $\sigma$ $\approx $ 0.06 mag level.  Consequently,  the difference in $m_{i}$ values between HE~1249--3121 and HE~2139--5432 is less than 1$\sigma$ and the difference from their nominal `expected' location in the metallicity-sensitive diagram is less than 2$\sigma$. 

The effect of the contribution of enhanced CH-features in the spectrum of a CEMP star to its location in the SkyMapper metallicity-sensitive diagram is undoubtedly a complex function of effective temperature, [Fe/H] and [C/Fe], as well as of [N/Fe] and [O/Fe].  \citet{JEN13b} have demonstrated that the CEMP-no stars (C-rich by definition) are also strongly enhanced in O, with typically [O/Fe] $\approx$ +1.6  dex.  Enhancements in N are also common among these stars with [N/Fe] $\approx$ +1.0 a typical value \citep[see Fig.\ 2 of][]{JEN13b}.  The role of enhanced N, in addition to enhanced C, is potentially significant for the SkyMapper EMP-candidate selection process as the violet CN-band head at $\lambda$3883\AA\/ is within the SkyMapper $v$-filter band pass.  Indeed Fig.\ 2 of \citet{Starkenburg17} illustrates the possible effect of enhanced C and N, though we note that their synthetic spectra are for T$_{\rm eff}$ = 4500 K, which corresponds to a $(g-i)_{0}$ colour exceeding 1.0 and is therefore outside our selection window.  The synthetic spectra are also calculated for [Fe/H] = \mbox{--3.0}, which enhances the effect and, although not explicitly stated, they have likely used [O/Fe] = +0.4 \citep{Starkenburg17}, a value which, for the C and N enhanced spectrum, is inconsistent with the results of \citet{JEN13b}. 

\begin{figure}
\centering
\includegraphics[angle=0.,width=0.48\textwidth]{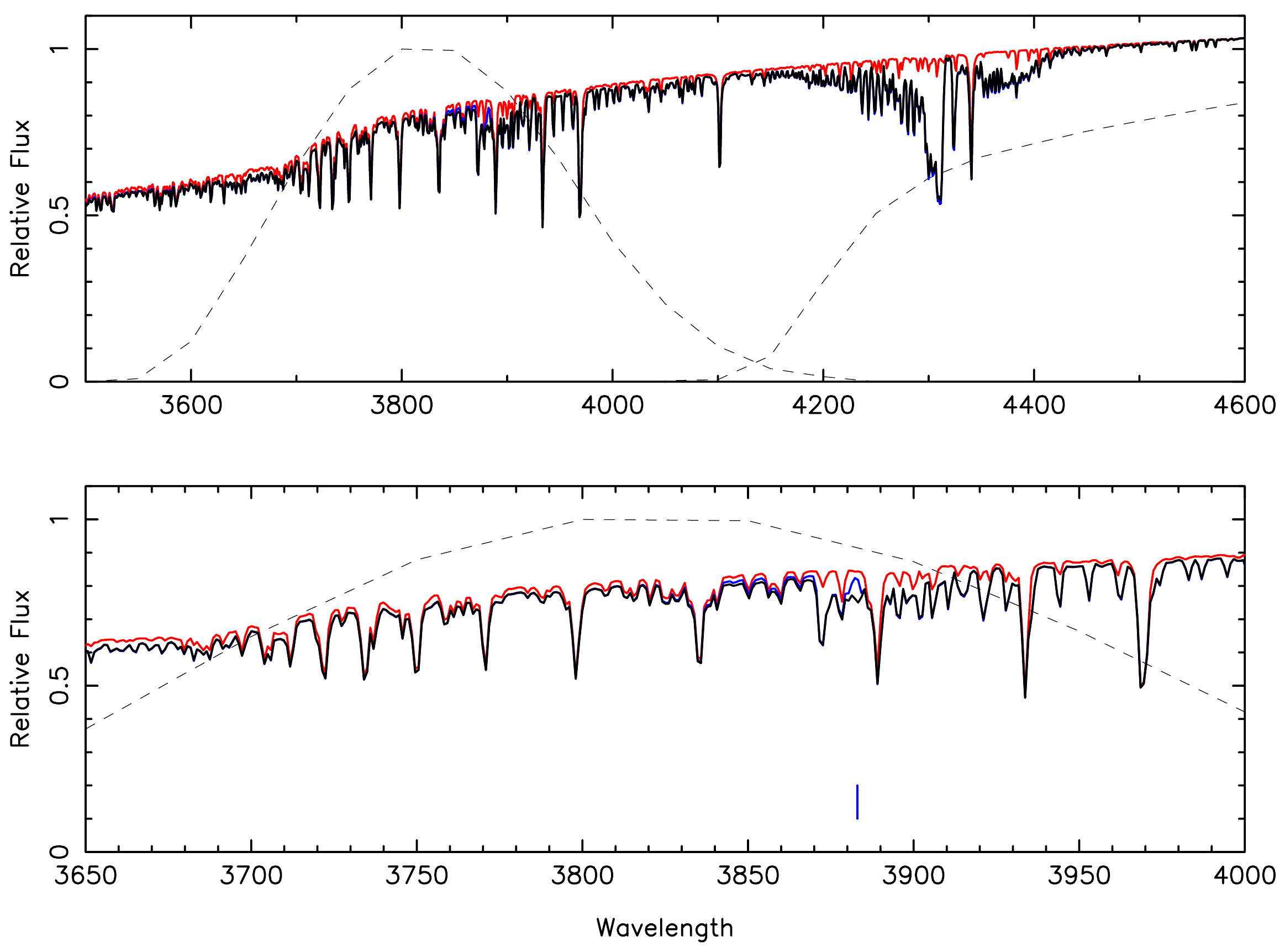}
\caption{Synthetic spectra, convolved to the 2.3m spectra resolution for $T_{\rm eff}$ = 5000 K, log $g$ = 1.9 and [Fe/H] = --4.0 dex.  The red line is for [C/Fe] = [N/Fe] = [O/Fe] = 0.0, the blue line is for [C/Fe] = +2.0 and [N/Fe] = [O/Fe] = 0.0 and the black line is for [C/Fe] = [N/Fe] = [O/Fe] = +2.0 dex.  The dashed lines show the bandpass of the SkyMapper $v$ filter and the blue part of the bandpass of the $g$ filter.  The lower panel focuses on the wavelengths included in the $v$-filter.  The vertical bar marks the $\lambda$3883\AA\/ CN-band head.}
\label{C-spec}
\end{figure}

As a qualitative illustration of the effects at a more relevant temperature and overall abundance, we show in Fig.\ \ref{C-spec} synthetic spectra, calculated as described in \S \ref{C_abund}, for $T_{\rm eff}$ = 5000 K ($(g-i)_{0}$ $\approx$ 0.65) at [Fe/H] = --4.0 and for the following specific cases: [C/Fe] = [N/Fe] = [O/Fe] = 0.0;  [C/Fe] = +2, [N/Fe] = [O/Fe] = 0.0; and, [C/Fe] = [N/Fe] = [O/Fe] = +2.0 dex.  Shown also in the panels are the bandpass of the SkyMapper $v$ filter and the blue section of the bandpass for the $g$ filter 
($g$ extends to $\sim$6600\AA) taken from \citet{MSB11}.  While there are clear differences in the strengths of the G-band, as shown in the lower panel there are also notable differences between the spectra in the vicinity of $\sim$3900\AA, although they are not large.  
The synthetic spectra with only C-enhanced and the spectra with C, N and O enhanced are very similar with only a very slight difference between the two visible in the vicinity of the $\lambda$3883\AA\/ CN-band.

\begin{figure}
\centering
\includegraphics[angle=0.,width=0.48\textwidth]{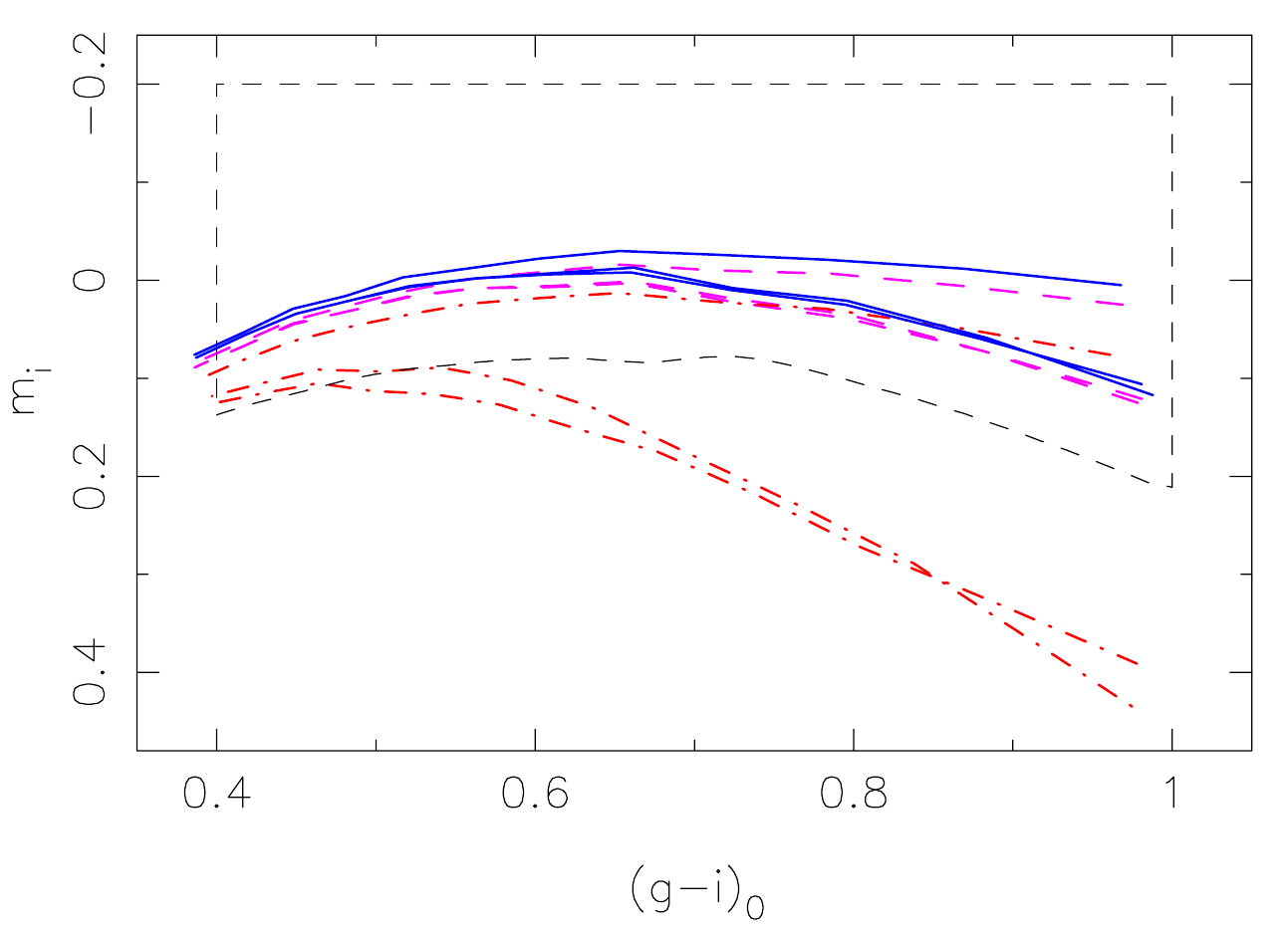}
\caption{Photometric indices $m_{i}$ and $(g-i)_{0}$ calculated from synthetic spectra.  The solid blue lines are for [Fe/H] = --5.0 and [C/Fe] = [N/Fe] = [O/Fe] = 0.0, [C/Fe] = +3.0 and [N/Fe] = [O/Fe] = 0.0, and [C/Fe] = [N/Fe] = [O/Fe] = +3 (the latter two are indistinguishable).
The magenta dashed lines are for [Fe/H] = --4.0 and [C/Fe] = [N/Fe] = [O/Fe] = 0.0, [C/Fe] = +2.0 and [N/Fe] = [O/Fe] = 0.0, and [C/Fe] = [N/Fe] = [O/Fe] = +2 (the latter two are again indistinguishable).  The red dot-dash lines are for [Fe/H] = --3.0 and [C/Fe] = [N/Fe] = [O/Fe] = 0.0, [C/Fe] = +2.0 and [N/Fe] = [O/Fe] = 0.0, and [C/Fe] = [N/Fe] = [O/Fe] = +2.  Shown also with dashed black lines is the adopted photometric selection window.  The horizontal scale is the same as for Fig.\ \ref{known_fig} but the range of the y-axis has been reduced.  }
\label{syn_phot}
\end{figure}

To investigate this further we show in Fig.\ \ref{syn_phot} a synthetic SkyMapper metallicity-sensitive diagram that we have generated by applying the SkyMapper filter bandpasses to the flux distributions of the synthetic spectra, calculated for a set of temperature, gravity, [Fe/H] and [C/Fe], [N/Fe] and [O/Fe] values.   \citet{Chiti2019} present similar calculations.  In each case the $T_{\rm eff}$ and log~$g$ values follow the giant branch isochrone appropriate for the metallicity.
While the normalization, particularly as regards the $m_{i}$ values, is not necessarily perfect, it is clearly reasonably consistent with the observations (see Figs.\ \ref{cmd_knownEMP} and \ref{known_fig}).  What is apparent from this figure is that the $m_{i}$ values are not strongly sensitive to enhancement in [C/Fe] alone, or in [C/Fe], [N/Fe] and [O/Fe], for [Fe/H] = --4.0 dex and lower. 
Specifically, Fig.\ \ref{syn_phot} shows that at [Fe/H] = --5 and --4, the curves for just C-enhanced, or C, N and O enhanced values are virtually indistinguishable and remain within the selection window.  Consequently, in the absence of any photometric uncertainties or systematic zero point errors, an unbiased selection of stars falling in the selection window should not exclude any C-rich (or CNO-rich) stars with [Fe/H] $\leq$ \mbox{--4.0} dex.  It appears, therefore, that whether Group~III stars with relatively low [Fe/H] values fall inside or outside the selection box (such as the HE~1249--3121 and HE~2139--5432 pair) may be more related to errors in the $v$-filter photometry than to changes in $v$-band flux resulting from the higher carbon (and higher C, N and O) abundance.

At [Fe/H] = --3.0, the curves for C-enhanced, and C, N and O enhanced abundances, drop out of the selection box even at relatively blue colours, and become more offset as the temperature decreases.  Consequently, in the absence of significant photometric uncertainties, an unbiased sample from within the selection box will likely lack, for [Fe/H] $\geq$ \mbox{--3.0} dex,  stars strongly enhanced in C (or strongly enhanced in C, N and O).  This is consistent with the fact that the Group~I stars (see Fig.\ \ref{ac_fig}) in the known-stars sample fall outside the selection window (see Fig\ \ref{known_fig}): they are C-rich (and likely also N and O-rich) and have [Fe/H] values at or above --3.0 dex.  While to do so is beyond the scope of this paper, it is likely that we could explain the specific location of these stars in Fig.\ \ref{known_fig} with synthetic photometry and appropriate choices of 
T$_{\rm eff}$, [Fe/H] and [C/Fe], [N/Fe] and [O/Fe].

What is clear from this analysis is that, in regard to Group~III stars, because of the overall uncertainty in the $v$-band photometry and the small number of stars involved, we cannot make any strong statements about completeness from the current set of [Fe/H]$_{fitter}$ values, which are derived from the low resolution spectra of stars within the photometric selection window.  We have revealed at least one new candidate (SMSS J160540.18--144323.1 with [Fe/H] $<$ --4.75 and [C/Fe] likely exceeding +2; (see \citet{TN18} for the actual values), but it is unclear how many Group~III stars at [Fe/H] $\approx$ --4 or thereabouts have been overlooked because, as Fig.\ \ref{syn_phot} shows, the strong CH-features do move the star downwards (particularly at redder colours) in the metallicity-sensitive diagram into the region where the fraction of candidates observed is low, and the `contamination' from somewhat more metal-rich objects is large.  

What we can say with some security is that we do not appear to be seriously incomplete as regards the discovery of carbon-normal stars (which includes Group~II objects), and that these dominate our sample.  Therefore, given that the three stars with 
[Fe/H]$_{fitter}$ $\leq$ --4.25 in our sample are all likely C-rich objects (see Figs \ref{Gband_fig} and \ref{Gband_fits}), we suggest that the apparent drop in the MDF in Fig.\ \ref{MDF_fig}  for [Fe/H] values below --4.0 likely applies principally to the carbon-normal (i.e., C scales with Fe) population.  Equally, our sample provides insufficient information to address the form of the MDF for the Group~III (CEMP-no) stars --- such stars may also exhibit the drop in the MDF or may be stochastically distributed in [Fe/H] to lower values.

\section{Discussion} \label{discuss-sect}

As a number of authors \citep[e.g.][]{BC05,JEN13b,FN15,NY19} have suggested, the existence of two populations of extremely metal-poor stars with different [C/Fe] values implies the existence of two different star-formation processes at early epochs and low [Fe/H] values.  Specifically, the CEMP-no stars, particularly those classified as Group~III where the C and Fe abundances are not strongly correlated, are thought to be objects formed via gas cooling processes that relied on high carbon  \citep[and probably oxygen, see, e.g.][]{JEN13b} abundances to achieve the necessary conditions for the formation of low-mass long-lived stars \citep[e.g.][]{BL03,AF07}.  By contrast, the extremely metal-poor stars where the C and Fe abundances tend to scale together, i.e., our carbon-normal category that includes the \citet{Yoon16} Group~II stars, are thought to be objects where dust cooling is more relevant to the formation of low-mass long-lived stars, i.e., a more `normal' star formation process \citep[e.g.][]{Schneider12,Chiaki17}.  Consequently, the Group~III stars are regarded as more `pristine' in that, in a given star formation environment, their formation may have preceded that of the carbon-normal stars.  Their element abundance ratios then provide constraints on the mass function of the Pop~III progenitors \citep[e.g.][]{FN15,TN17}.

Our principle contribution in this area is the demonstration of an apparent abrupt drop in the metallicity distribution function at 
[Fe/H]$_{fitter}$ $\approx$ --4.0 for a sample that we assert is relatively unbiased as regards metallicity and which is dominated by carbon-normal stars, i.e., those where C scales with Fe.
\citet{DY13}, using an entirely independent sample of EMP-stars, found a turnover in the MDF at a similar metallicity though that work did not explicitly connect the MDF-turnover with the carbon-normal population.  Further, as is evident in Fig.\ \ref{ac_fig}, the 155 carbon-normal objects (i.e.,  those with [C/Fe]$_{1D, LTE}$ $\leq$ 1.0) in our sample of known abundance stars, also show a turnover in the metallicity distribution at [Fe/H]$_{1D, LTE}$ $\approx$ --4.2 dex, consistent with that seen in Fig.\ \ref{MDF_fig}.  Further, \citet{Yoon18} using a different sample, have asserted that CEMP-no stars, primarily Group~III objects, dominate at 
metallicities [Fe/H]$_{1D, LTE}$ $\approx$ --4.0 and lower, though the number of objects in their sample at the lowest metallicities is small.  \cite{NY19} discuss these topics in detail, with a particular emphasis on correcting the derived 1D, LTE carbon and iron abundances for 3D and NLTE effects, which are significant 
\citep{MA05,RC06,RC18}.   They nevertheless find, using a sample that is dominated by that of \citet{DY13}, that the drop in the relative frequency in carbon-normal stars remains although the location moves from [Fe/H]$_{1D,LTE}$ $\approx$ --4.2 to [Fe/H]$_{3D, NLTE}$ $\approx$ --4.0 \citep{NY19}.  Consistent with these earlier results, we propose that a metallicity of this order marks, within a specific environment, the transition from dominant carbon (and oxygen) driven gas cooling processes for low-mass star formation to dominant more normal dust cooling driven star formation.

Carbon-normal, or indeed Group II, stars appear to be extremely rare for [Fe/H]$_{1D,LTE}$ $<$ --4.2 dex.  The most well-known example is SDSS J102915+172927 \citep{Caffau11}, which, with [Fe/H]$_{1D, LTE}$ $\approx$ --4.73 and [C/H]$_{1D, LTE}$ $<$ 
--3.8, has [C/Fe]$_{1D, LTE}$ $<$ 0.9 and A(C)$_{1D, LTE}$ $<$ 4.6 dex.  It can therefore arguably be classified as a `carbon-normal' object.  More recently, \citet{ES18} have reported abundances for the star Pristine\_221.8781+9.7844 finding [Fe/H]$_{1D,LTE}$ = --4.66 and [C/Fe]$_{1D, LTE}$ $<$ 1.76 and A(C)$_{1D, LTE}$ $<$ 5.6 \citep{ES18}.  It is not clear how to classify this star -- a carbon abundance at least 0.8 dex lower is required if the star is `carbon-normal', while a value $\sim$0.3 dex lower would make the star similar to HE~0557--4840\footnote{HE~0557--4840 is in the SkyMapper DR1.1 database but it did not make it into our list of known stars meeting the photometric selection criteria, i.e., the stars plotted in Fig.\ \ref{ac_fig} for example, as with a value of 2, its database entry fails the {\it nch\_max} = 1 selection criteria.  The listed photometry, however, yields $(g-i)_{0}$ = 0.61 and $m_{i}$ = --0.06 placing it well within the photometric selection window in Fig.\ \ref{known_fig}.  It would then be plotted as a filled-star symbol in Fig.\ \ref{ac_fig}.} that has a measured [Fe/H]$_{1D,LTE}$ = \mbox{--4.75}, [C/Fe]$_{1D, LTE}$ = 1.66 and A(C)$_{1D, LTE}$ = 5.3 \citep{Norris07}.  This object is perhaps best classified as the most carbon-poor of the Group III stars, and Pristine\_221.8781+9.7844 may well be similar.

\section{Summary and Outlook}

We have described here an extensive and on-going program that uses a combination of SkyMapper DR1.1 photometry and low-resolution spectroscopic follow-up with the ANU 2.3m telescope to identify extremely metal-poor stars in the Galactic halo.  By deliberate choice of the colour range for the candidate stars, the survey targets EMP giants.  It has proved to be a very efficient approach with over 40\% of the photometric candidates having [Fe/H] $\leq$ --2.75 as derived from the low-resolution spectra.  The low-resolution spectra allow the identification of 142 stars within the selection window boundaries that have 
[Fe/H]$_{fitter}$ $<$ --3.0 and which, to our knowledge, have not been previously studied.  The parameters for these stars are given in Table \ref{tab:data} while Table \ref{tab:data2} gives the same data for a further 29 candidates whose DR1.1 photometry falls outside the selection window or which fail the adopted  DR1.1 photometric selection criteria.

The majority of these EMP candidates have been observed at high resolution with the MIKE echelle spectrograph on the Magellan Clay 6.5m telescope and the abundance analyses will be presented in a forthcoming paper \citep{DY18}.  When combined with the high-dispersion follow-up of SkyMapper commissioning-era EMP candidates discussed in \citet{HJ15} and \citet{AFM19}, the results will represent by far the largest sample of consistently selected and analysed element abundance measurements for extremely metal-poor stars.  The most metal-poor star discovered in this survey to date is SMSS J160540.18--144323.1 which has [Fe/H]$_{fitter}$ $<$ --4.75 and [C/Fe] likely exceeding +2 dex.  A detailed discussion of the abundances for this star, again based on Magellan/MIKE spectra, is given in \citet{TN18} where it is shown that the star has [Fe/H]$_{1D,LTE}$ = --6.2 $\pm$ 0.2 and [C/Fe]$_{1D,LTE}$ = 3.9 
$\pm$ 0.2 dex.

We have also demonstrated, using metal-poor stars with known abundances that meet our DR1.1 selection criteria together with spectrum synthesis applied to the low-resolution spectra of the more metal-poor stars, that the observed sample is dominated by `carbon-normal' stars, i.e., those stars in which the carbon abundance scales with iron.  Stars with very strong carbon enhancements, particularly CEMP-stars with neutron-capture element enhancements and [Fe/H] $\geq$ --3.3 (approximately), are likely to be entirely absent from the current photometric selection due to the reduced flux, relative to a carbon-normal star, in the bandpass of the $v$-filter.  Nevertheless, CEMP-stars (likely mostly CEMP-no) are present in the current set of stars with 2.3m telescope spectroscopy, but the level of completeness for such stars is a complex function of [Fe/H], [C/Fe] and $T_{\rm eff}$ and uncertainties in the $v$-band photometry which we have not attempted to quantify in detail.

On the other hand, the sample of carbon-normal stars (i.e, those where the carbon and iron abundances scale together) with low-resolution spectra, which dominate the numbers, is largely unbiased with respect to metallicity, and the metallicity distribution function for these stars appears to drop abruptly for [Fe/H]$_{fitter}$ $\approx$ --4.2 dex.  We argue, as have others, that this limit represents, within a specific environment, the threshold abundance for the formation of low-mass stars via dust-driven gas-cooling process analogous to those of the present-day.   In that sense these stars are `less primordial' than the carbon-rich CEMP-no stars (carbon and iron abundances are uncorrelated) for which the low-mass star formation at the earliest epochs is governed by gas-cooling processes involving carbon and oxygen.  Unfortunately, the current sample does not provide any significant constraints on the MDF for these latter stars. 

As for the future of the SkyMapper search for EMP stars, it is clear that the current selection process is largely constrained by the  (systematic and photometric) errors in the $v$-filter magnitudes in the DR1.1 photometry database.  The application of the $m_{i}$ index is clearly very useful in determining metal-poor candidates that have [Fe/H] $\leq$ --2.0, but the level of further metallicity discrimination at lower abundances is not strong with the DR1.1 photometry.  The SkyMapper DR2 data release \citep{Onken2019} is likely to be a significant improvement over the DR1.1 photometry for two reasons.  First, since DR2 contains the first main survey data, which goes substantially fainter than the shallow survey underlying DR1.1, there should be significant improvements in the photometric precision of the $v$-magnitudes for the $g$ $<$ 16 stars with main survey photometry.  Second, as described in \citet{Onken2019}, the calibration of the zeropoint of the DR2 photometry across the sky is expected to be much more uniform than for DR1.1 as it is  based on Gaia DR2 magnitudes.  Both these improvements should lead to enhanced metallicity discrimination with the DR2 photometry.

Regarding the spectroscopic follow-up, the 2.3m telescope program of low-resolution spectroscopy will continue, aiming primarily at the best DR2 EMP-candidates.  We also have ancillary science program status with the forthcoming Taipan survey \citep{EdC17} in which the allocation of ``spare fibres'' to EMP-candiates will allow low-resolution spectra to be obtained for stars with 15 $\leq$ $g$ 
$\leq$ 16 on an on-going basis, using the multi-fibre spectroscopic system at the UK Schmidt telecope.  The EMP-candidates to be observed in the Taipan survey, typically a few per 6-degree diameter survey field, are likely to provide a largely unbiased sampling of the candidate list, and over the $\sim$4 year duration of the Taipan survey, we expect to obtain 10,000 or more EMP-candidate spectra.
Further, since our selection process is strongly giant focussed, we also intend to make use of the astrometric information provided in the Gaia DR2 database \citep{GaiaDR2} to remove from the candidate list of any contaminating dwarfs that masquerade as metal-poor giants, such as young stars with Ca {\sc ii} H and K emission that increases the $v$-filter flux.  Use of Gaia data will also allow us to extend the selection to include somewhat cooler giants than the current cutoff at $(g-i)_{0}$ = 1.0 ($T_{\rm eff}$ $\approx$ 4500 K).  Such luminous stars are rare but probe a large spatial volume. The improved DR2 photometry will also make it worthwhile to explore additional selection information that incorporates other SkyMapper photometry beyond just $vgi$, as well as near- and mid-IR photometry \citep[e.g.][]{SC14}.  It may then be possible, for example,  to specifically identify and target candidate CEMP-no stars.  The quest for evermore primordial stars will continue.

\section*{Acknowledgements}

The authors would like to acknowledge the many people, both at Siding Spring Observatory and at Mt Stromlo, who have contributed to the SkyMapper Southern Sky Survey through the on-going operation of the telescope and its systems, the processing of the imaging, and the generation and distribution of the data products.  Without these contributions there would be no EMP-star project.

The authors also acknowledge and appreciate the contribution of Aaron Dotter who calculated isochrones consistent with those of the Dartmouth Stellar Evolution Database but which have lower metallicities.

SkyMapper research on EMP stars has been supported in part though the Australian Research Council (ARC)
Discovery Projects DP120101237 and DP150103294 (Lead-CI Da~Costa).  A.~D.~M.\ is supported by an ARC Future Fellowship (FT160100206) and A.~R.~C.\ is supported in part by ARC Discovery Project DP160100637.  Parts of this research were conducted under the auspices of the ARC Centre of Excellence for All Sky Astrophysics in 3 Dimensions (ASTRO 3D), which is supported through project number CE170100013.
The project has also received funding 
from the European UnionÕs Horizon 2020 research and innovation programme under the Marie Sk\l odowska-Curie Grant Agreement 
No.\ 797100 (beneficiary: A.~F.~M.)


The national facility capability for SkyMapper has been funded through ARC LIEF grant LE130100104 from the Australian Research Council, awarded to the University of Sydney, the Australian National University, Swinburne University of Technology, the University of Queensland, the University of Western Australia, the University of Melbourne, Curtin University of Technology, Monash University and the Australian Astronomical Observatory. SkyMapper is owned and operated by The Australian National University's Research School of Astronomy and Astrophysics. The survey data were processed and provided by the SkyMapper Team at ANU\@. The SkyMapper node of the All-Sky Virtual Observatory (ASVO) is hosted at the National Computational Infrastructure (NCI).  Development and support the SkyMapper node of the ASVO has been funded in part by Astronomy Australia Limited (AAL) and the Australian Government through the Commonwealth's Education Investment Fund (EIF) and National Collaborative Research Infrastructure Strategy (NCRIS), particularly the National eResearch Collaboration Tools and Resources (NeCTAR) and the Australian National Data Service Projects (ANDS).

This work has made use of data from the European Space Agency (ESA) mission {\it Gaia} (https://www.cosmos.esa.int/gaia), processed by the 
{\it Gaia} Data Processing and Analysis Consortium (DPAC, https://www.cosmos.esa.int/web/gaia/dpac/consortium). Funding for the DPAC
has been provided by national institutions, in particular the institutions participating in the {\it Gaia} Multilateral Agreement.

We also acknowledge the traditional owners of the land on which the SkyMapper telescope stands, the Gamilaraay people, and pay our respects to elders past, present and emerging.

\end{document}